\documentclass[5p]{elsarticle}
\biboptions{sort&compress}
\usepackage{hyperref}
\hypersetup{
colorlinks = true,
linkcolor = blue,
anchorcolor = blue,
citecolor = blue,
filecolor = blue,
urlcolor = blue
}

\usepackage[utf8]{inputenc}
\usepackage{xcolor}
\usepackage{comment}
\usepackage{amsmath}
\usepackage{amssymb}
\usepackage{float}
\usepackage{multirow}
\usepackage[colorinlistoftodos]{todonotes}
\usepackage{lipsum}
\usepackage[nameinlink,capitalize]{cleveref}
\usepackage{placeins}

\usepackage{siunitx}
\usepackage[nolist]{acronym}
\usepackage{caption}
\usepackage{amssymb,amsmath,amsthm,enumitem}
\usepackage{gensymb}
\usepackage{subfigure}
\usepackage{wrapfig}
\usepackage{lscape}
\usepackage{rotating}
\usepackage[normalem]{ulem}
\useunder{\uline}{\ul}{}
\journal{Journal}
\DeclareUnicodeCharacter{FFFD}{fi}
\begin{document}
%\modulolinenumbers[0]

\begin{frontmatter}

\title{{\bf \LARGE Rapid quantification of dynamic and spall strength of metals across strain rates}}

%Suhas, Pinkesh Malhotra, Xingsheng, Taiki Nakata, John Fite, Justin Moreno, Fanuel Mammo, Stephanie Hernandez, Matt Shaffer, KT Ramesh, Tim Weihs

\author[a,b,c,d]{Suhas Eswarappa Prameela\corref{corrauthor}}
\ead{suhasep@mit.edu}
%\fnref{myfootnote}}
\author[e,m]{Christopher C. Walker}
\author[d,f,m]{Christopher S. DiMarco}
\author[d,g]{Debjoy D. Mallick}
\author[h]{Xingsheng Sun}
\author[i]{Stephanie Hernandez}
\author[j,k]{Taisuke Sasaki}
\author[l]{Justin W. Wilkerson}
\author[d,i]{K.T. Ramesh}
\author[e]{George M. Pharr}
\author[c,d]{Timothy P. Weihs\corref{corrauthor}}
\ead{weihs@jhu.edu}

\cortext[corrauthor]{Corresponding authors}

\address[a]{Department of Materials Science and Engineering, MIT, Cambridge, MA, USA}
\address[b]{Department of Aeronautics and Astronautics, MIT, Cambridge, MA, USA}
\address[c]{Department of Materials Science and Engineering, Johns Hopkins University, Baltimore, MD 21218, USA}
\address[d]{Hopkins Extreme Materials Institute, Johns Hopkins University, Baltimore, MD, 21218, USA}
\address[e]{Department of Materials Science and Engineering, Texas A\&M University, College Station, TX, 77843, USA}
\address[f]{Sindri Materials Corp., West Chester, PA, 19382, USA}
\address[g]{DEVCOM Army Research Laboratory, 321 Colleran Road, Aberdeen Proving Ground, MD, 21005-5066, USA}
\address[h]{Department of Mechanical and Aerospace Engineering, University of Kentucky, Lexington, KY 40506, USA}
\address[i]{Department of Mechanical Engineering, Johns Hopkins University, Baltimore, MD 21218, USA}
\address[j]{National Institute for Materials Science, Tsukuba 305-0047, Japan}
\address[k]{Center for Elements Strategy Initiative for Structural Materials (ESISM), Kyoto University, Yoshida Honmachi, Sakyo, Kyoto 606-8501, Japan}
\address[l]{Department of Mechanical Engineering, Texas A{\&}M University, College Station, TX, 77843, USA}
\address[m]{These authors contributed equally to this manuscript}

\begin{abstract}Over the last few decades, numerous small-scale mechanical testing techniques have been developed to evaluate the fundamental deformation mechanisms and failure behavior of metals. However, many of these techniques are insufficient to accurately predict material behavior at extreme strain rates, critical for applications such as car crashes, meteoroid impact on satellites, launch and re-entry of rockets, and protection materials to stop bullets. The response of metals and their microstructures under extreme dynamic conditions can be markedly different from that under quasistatic conditions. Traditionally, high strain rates and shock stresses are measured using cumbersome and expensive methods such as the Kolsky bar or large spall experiments. These methods are low throughput and do not facilitate high-fidelity microstructure-property linkages. In this work, we combine two powerful small-scale testing methods, custom nanoindentation, and laser-driven micro-flyer shock, to measure the dynamic and spall strength of metals. The nanoindentation system is configured to test samples from quasistatic to dynamic strain rate regimes (10$^{-3}$ s$^{-1}$ to 10$^{+4}$ s$^{-1}$). The laser-driven micro-flyer shock system can test samples through impact loading between 10$^{+5}$ s$^{-1}$ to 10$^{+7}$ s$^{-1}$ strain rates, triggering spall failure. The model material used for testing is Magnesium alloys, which are lightweight, possess high-specific strengths and have historically been challenging to design and strengthen due to their mechanical anisotropy. Here, we modulate their microstructure by adding or removing precipitates to demonstrate interesting upticks in strain rate sensitivity and evolution of dynamic strength. At high shock loading rates, we unravel an interesting paradigm where the spall strength of these materials converges, but the failure mechanisms are markedly different. Peak aging, considered to be a standard method to strengthen metallic alloys, causes catastrophic failure, faring much worse than solutionized alloys. Our high throughput testing framework not only quantifies strength but also teases out unexplored failure mechanisms at extreme strain rates, providing valuable insights for the rapid design and improvement of metals for extreme environments.
  
\end{abstract}{}


\end{frontmatter}

\tableofcontents

\section{Introduction}

Traditionally, the mechanical properties of bulk structural materials across strain rate regimes have been probed using bulk techniques. For the quasistatic strain rate regime, this evaluation has been done through tension, compression, bending, and torsion experiments. For the dynamic strain rate regime, researchers have employed plate impact, shock, isentropic compression, and Kolsky bar experiments \cite{meyers1994dynamic}. These test protocols are useful for bulk samples but pose challenges when rapid screening of properties is needed for materials design and discovery for extreme environments \cite{eswarappa2023materials}. Furthermore, some large-scale shock/high strain rate experiments can be highly destructive, expensive, and logistically burdensome to implement. There have been several efforts to implement small-scale mechanical testing methods, both at the microscale and nanoscale. Micro-tensile, micro-cantilever, and micro-pillar tests and their counterparts at the nanoscale have allowed researchers to carry out site-specific or volume-specific experiments and obtain the attendant mechanical response \cite{weihs1988mechanical,rajagopalan2019microelectromechanical,nix2011micro}. These experiments have also been coupled with various diagnostic tools such as scanning electron microscopy (SEM), transmission electron microscopy (TEM), and other X-ray or beamline instruments that offer insights into deformation mechanisms and in-situ tracking of key parameters such as texture/precipitate evolution, extent of slip behavior, and twin volume fractions \cite{barnard2017synchrotron,van2017miniaturized} in metals.  \\

Over the last two decades, several efforts have pushed small-scale mechanical testing protocols to adopt testing conditions to recapitulate those encountered in extreme environments. For example, there are several active efforts to push the popular nanoindentation technique to high strain rate regimes often encountered during car crashes and ballistic impact. These efforts have focused on improving the calibration methods, measurement strategies, noise reduction techniques, and expanding the range of material systems that can be tested \cite{guillonneau2018nanomechanical,zehnder2018high, somekawa2012high, merle2019critical,merle2020extending}. Some studies have looked at the impact of continuous stiffness measurement methods (CSM) on hardness overestimation and subsequent challenges in measuring strain rate sensitivity \cite{sudharshan2017ultra}. Managing noise levels and data analysis become especially challenging at high strain rates. Advanced sensors and high-frequency modulation techniques are being developed to circumvent these challenges. \\

There have also been several efforts to mimic shock-loading conditions through launched particles and plates. These impact experiments cause materials to experience deformation at high strain rates. For example, Laser-Induced Particle Impact Test (LIPIT) experiments have been successfully implemented to test various metallic alloys, ceramics, and other structural materials \cite{imbriglio2019adhesion, tiamiyu2022nanotwinning}. In these experiments, a laser is used to accelerate micro-particles ($\sim$10-50 $\mu$m) at varying speeds ($\sim$100 ms$^{-1}$ to 900 ms$^{-1}$) towards a target material. These experiments have helped to elucidate adhesion mechanisms, cold spray mechanisms, recrystallization, plasticity, and damage at extreme strain rates.  Similarly, laser energy can be used to accelerate thin circular metal disks to mimic plate impact and interrogate spall behavior. Several recent studies have looked at laser-driven micro-flyer shock experiments for testing various metallic alloys, single crystals, ceramic carbides, and other structural materials \cite{mallick2019laser, mallick2019shock, mallick2021spall}.  \\

%The dynamic strength of Mg and Mg alloys, typically in the strain rate regime of ($\sim10^{3-4}$ s$^{-1}$), has also been investigated, particularly in the last decade through various collaborative efforts and consortia. Well-developed Kolsky bar experiments (tension, compression, torsion, hybrid modes) have helped unravel several interesting microstructure property linkages \cite{wei2021insights,bhattacharyya2016crystal}. For example, texture weakening has been observed to help minimize the plastic anisotropy and extension twinning in high strain rate (compression mode) experiments \cite{asgari2014texture,asgari2015grain}. While several studies focus on the influence of texture, grain size, and chemistry, there are still unclear linkages between precipitate density or type and the dynamic strength of Mg alloys. High-throughput experiments can accelerate testing of important microstructures such as precipitates and help characterize their dynamic performance. 

In this study, we chose a Magnesium (Mg) alloy as the model material to demonstrate rapid quantification of strength across various strain rates and link them to attendant microstructures and plasticity mechanisms. Our testing framework (Fig.~\ref{fig:1}) employs custom nanoindentation and laser-driven micro-flyer shock experiments to help quantify the quasistatic, dynamic, and spall strength of these metallic alloys. To explore the effect of heterogeneous inclusions, such as precipitates, we apply similar testing protocols on two different variants of the same metallic alloy, peak-aged and solutionized samples with and without precipitates, respectively. High-throughput experiments can accelerate the testing of these variants while shedding insights into how microstructural features such as precipitates that are conventionally found to be favorable for metal strengthening at quasistatic strain rates behave in dynamic and spall regimes.  \\
\\

When Mg alloys are deformed at high strain rates ($\sim10^{4-8}$ s$^{-1}$) through impact experiments, shockwaves are generated that first load the material in uniaxial compression. When these shock waves meet free surfaces, they reflect as rarefaction fans that can intersect within the material, resulting in dynamic tensile loading that is nearly hydrostatic. This dynamic hydrostatic tension causes the material to fail through a process called spallation, with such spall failures typically driven in metals by void nucleation, growth, and coalescence. The high specific strength of Mg alloys offers a compelling reason to pursue directions related to enhancing spall strength \cite{mallick2020brief}. The spall strength, defined as resistance to spallation, strongly depends on microstructural features such as grain size, texture, precipitate type, and volume fraction. The spall strength is also a strong function of the tensile strain rate and potentially of the degree of shock compression before the tensile loading occurs. While there have been studies on the dynamic and spall failure of Mg single crystals and alloys \cite{mallick2021spall,yu2017influence,de2017spall,hazell2012influence,farbaniec2017spall,sun2022uncertainty} there is still a lack of clarity on the microstructure property linkages. Recent spall studies in Mg alloys have shown the importance of precipitate size and distribution. In the case of AZ31 alloy, some large-size precipitates result in catalytic void nucleation and accelerated spall failure in the materials \cite{williams2020concise,farbaniec2017spall}. Taken together, these prior studies make a strong case to employ high-throughput techniques to quickly quantify spall strength and to use that data along with the quasistatic and dynamic strengths of Mg alloys to understand the microstructure-property linkages in these materials. 

\section{Materials and Methods}
\subsection{Thermomechanical Processing and Sample Preparation}
\vspace{3mm}
The Mg-5Zn (at\%) alloy, referred to as Z5, was prepared by melting and mixing high purity Mg (99.97\%, Grade II, U.S. Magnesium) and Zn (99.999\%, Alfa Aesar) in an argon atmosphere and casting into bars. The bars were solutionized at $500 ^{\circ}$C for 25 hours to remove any preexisting precipitates from the casting process. The bars were then cut into rectangular pieces of 20 mm by 10 mm by 5 mm. Some of these pieces were then peak-aged at $150 ^{\circ}$C for 99 hours to disperse precipitates throughout the bulk. \\

The solutionized and peak-aged Z5 pieces were then mechanically polished for further characterization and testing. A field emission scanning electron microscope (SEM, Carl Zeiss Crossbeam 1540 EsB FIB/SEM) was used for electron microscopy studies and electron backscattered diffraction(EBSD) scans. The SEM has an HKL EBSD system with the Channel 5 software for EBSD analysis. For scanning transmission electron microscope(STEM) studies, thin foil specimens, including grain boundaries, were placed by a standard lift-out technique using a dual beam FIB/SEM, FEI Helios G4. % the samples were polished down to 100 um thick foils, followed by twin jet polishing to create holes and PIPS (Precision Ion Polishing System) to clean the boundary of the hole to make it electron transparent and ready for observations. 
An FEI Titan G2 80–200 STEM was used for annular dark\textendash field (ADF) imaging and diffraction studies. 
%%%%%%%%%%%%%SUB-SECTION%%%%%%%%%%%%%%%%%%%%%%%%%%%%%%%%%%%%%
\subsection{Low to High Strain Rate Nanoindentation Experiments} 
\vspace{3mm}

\begin{acronym}\itemsep-30pt
	\acro{ISE}{indentation size effect}
	\acro{CSM}{continuous stiffness measurements}
	\acro{RMS}{root means square}
	
	\acro{h}[$h$]{indentation depth}
	\acro{hc}[$h_c$]{contact depth}
	\acro{hch}[$h_c/h$]{ratio of contact depth to depth}
	
	\acro{hdot}[$\dot h$]{velocity}
	
	\acro{isr}[$\dot \varepsilon_i$]{indentation strain rate}
	
	\acro{P}[$P$]{load}
	\acro{Pdot}[$\dot P$]{loading rate}
	\acro{PdotP}[$\dot P/P$]{ratio of loading rate to load}
	
	\acro{Ac}[$A_c$]{contact area}
	
	\acro{H}[$H$]{hardness}
	\acro{Hdot}[$\dot H$]{change in hardness over time}
	\acro{E}[$E$]{young's modulus}
	
\end{acronym}	
	
Nanoindentation is a well-established method for measuring the basic mechanical properties of materials and provides many advantages to traditional hardness testing \cite{OliverPharr:1992, OliverPharr:2004}. In this work, nanoindentation was performed using a custom nanoindentation system made by KLA (USA) that is capable of testing in both the quasistatic and dynamic strain rate regimes (details of the custom instrumentation are outlined by Hackett et al. \cite{HackettHSRMethods:2023}). A diamond indenter tip with the three-sided Berkovich geometry was used for all indents. The schematic of the nanoindentation system equipped with both the quasistatic straining protocol and dynamic straining protocol is shown in (Fig.~\ref{fig:1} A and B). For the quasistatic indents, a constant strain rate was achieved by loading such that the \ac{PdotP} was constant. With a constant \ac{PdotP}, the \ac{isr} can be calculated as

\begin{equation}\label{isr_eq}
	\epsilon_i \equiv \frac{\dot{h}}{h} = \frac{1}{2}\left(\frac{\dot{P}}{P} - \frac{\dot{H}}{H}\right)
\end{equation}

\noindent with \ac{hdot}, \ac{h}, \ac{Pdot}, \ac{P}, \ac{Hdot}, and \ac{H} \cite{LucasOliver:1997}. When \ac{H} is constant, $\dot{H} = 0$ and equation \eqref{isr_eq} simplifies to

\begin{equation}\label{isr_eq_simple}
	\epsilon_i \equiv \dot{h}/h = \frac{\dot{P}}{2P}.
\end{equation}

\noindent \Acl{H} is constant as a function of depth for the tested Mg alloys, so equation \eqref{isr_eq_simple} can be used. Quasi-static indents were performed to a maximum load of \SI{200}{\milli\newton} and at \acp{PdotP} of \SIlist{0.02;0.2;2.0}{\per\second}. During loading, \ac{CSM} were taken by applying a \SI{110}{\hertz} oscillation to the load such that the \ac{RMS} amplitude of the oscillation was 10\% of the load as prescribed by Phani et al. \cite{PhaniMethod:2020}. Results for \ac{H} and \ac{isr} were averaged over an \acl{h} of \SIrange{500}{2500}{\nano\meter}. A linear least squares fit of \acl{hc} vs. \acl{h} was used to determine a value for the \ac{hch} in each material. This fit results in an \ac{hch} of 0.94 and 0.98 for the solutionized and peak-aged material, respectively. Optical images of each indent were taken to confirm the residual contact impression matched the results of the \ac{CSM} measurements.\\

An impact testing approach was used for the high strain rate indentation,  as described by Phani et al. \cite{PhaniHSRMath:2023} and Hackett et al. \cite{HackettHSRMethods:2023}. To account for the dynamic overload during impact, loads of \SIlist{30;50}{\milli\newton} were used so the final load on the sample would be between \SIrange{200}{300}{\milli\newton}, reaching comparable indent sizes with the quasistatic testing. Due to the nature of the impact test, a constant \ac{PdotP} can not be maintained. Thus \ac{isr} during these impact tests is not constant, allowing a range of strain rates to be reported from each test. Additionally, \ac{CSM} cannot be used to measure \ac{H} as \ac{CSM} has been shown to have problems at high strain rates \cite{Merle:2019}, and the instrument cannot apply the oscillation required for \ac{CSM} at a frequency suitable for the $<$ \SI{1}{\milli\second} impact test. Without \ac{CSM}, \acl{H} is calculated with

\begin{align}
	h_c &= \frac{h_c}{h} \times h\\
	A_c &= \sum_{n=0}^{5} C_n h_c^{2^{(1-n)}}\\
	H &= P/A_c
\end{align}

\noindent where \ac{hch} was first determined during the quasistatic testing, and $C_n$ are a series of constants that describe the tip shape \cite{OliverPharr:1992, HackettHSRMethods:2023}. However, optical imaging of the residual contact impressions showed that the calculated \ac{Ac} when using \ac{hch} from quasistatic testing did not align with the physical size of the indent. A new \ac{hch} was calculated, \numlist{0.90;1.0}, for solutionized and peak-aged, respectively, so that \ac{Ac} aligned with the measured contact area for the dynamic indents. A miniature ICP force sensor from PCB Piezoelectronics (USA) was added to the system to measure the applied load during dynamic testing. The addition of the piezoelectric force sensor lowers the instrument frame stiffness from \SI{25}{\mega\newton/\meter} to \SI{8}{\mega\newton/\meter}, but a standard frame stiffness correction can account for this correctly\cite{OliverPharr:1992}. Hardness and \acl{isr} were binned into 120 half-open bins between strain rates of \SI{e1}{\per\second} and \SI{e4}{\per\second}.
%%%%%%%%%%%%%SUB-SECTION%%%%%%%%%%%%%%%%%%%%%%%%%%%%%%%%%%%%%
\subsection{Laser Driven Micro-Flyer Shock Experiments} 
\vspace{3mm}
Laser shock methods are well-poised for high-throughput dynamic experiments; in contrast to conventional methods, these methods attain similar energy densities while operating more safely, at smaller scales requiring significantly less material, and at much lower overall expense \citep{paisley1991laser, paisley2007experimental,brown2012simplified, mallick2019laser, li2022high}.  The Laser-Driven Micro-Flyer (LDMF) Shock experimental set-up, as shown in (Fig.~\ref{fig:1}) is a subset of these laser shock methods that utilizes the energy from a laser pulse to accelerate a micro-scale flyer plate to achieve a high-velocity impact with a target, achieving tensile strain rates of $\ge\mathcal{O}$(10$^6$) s$^{-1}$ during spall. We determine the dynamic response of the material through Photonic Doppler Velocimetry (PDV) measurements on the target’s free surface during the impact \citep{mallick2019laser, dolan2020extreme}. In contrast to traditional plate impact experiments, which average $\sim$ 1-3 experiments per day, LDMFs can obtain similar impact energy densities and can easily exceed 100 experiments per day. While the small scale of the LDMF experiment significantly reduces material waste, it requires careful consideration of the microstructure and deformation length scales at play. %Even though these laser-driven micro-flyer shock methods have been around for several decades, there has only recently been increased attention towards high-throughput experiments with these techniques. 
Even with recent attention to several shock compression applications \citep{li2022high}, very few investigations optimized to study spall failure. Recently, we have developed an LDMF system and methodology for investigating spall failure with high-throughput while maintaining reasonably high fidelity. In this effort, there are two critical challenges for establishing confidence and reproducibility: (1) maintaining a high degree of flyer planarity from launch to impact, and (2) establishing conditions such that a consistently high-quality PDV signal is obtained. This work is among the first experimental demonstrations \cite{dimarco2023} of our new methods to address these challenges. \\

The LDMF experiment consists of three sections: (1) the pulse laser and free-space optics; (2) the launch package; and (3) the PDV diagnostics (Fig.~\ref{fig:1} E). The goal of the first section is to emit and manipulate the energetic, spatial, and temporal characteristics of the driving laser pulse to achieve optimal launch conditions (i.e., planar flyer at a desired velocity).  Our system uses a 1064 nm Nd: YAG 2.5 J 10 Hz 10 ns Spectra-Physics Quanta-Ray 350 with a beam quality (M2 value) of ~15.  The pulse energy drives the flyer velocity through the laser fluence.  The pulse duration must be sufficiently long relative to the round-trip shock wave time inside the flyer to prevent reverberations during the launch that can break up the flyer \citep{brown2012simplified}. We employ an optical cavity to lengthen the pulse duration from $\sim$10 ns to $\sim$21 ns, which is sufficient for the flyer material and thickness used in these studies.  Lastly, and perhaps most critical, the acceleration of a planar and intact flyer requires a homogenized beam profile, and so the beam is shaped just before the flyer launch. We utilize a custom-built Diffractive Optical Element (DOE) by Silios Technologies to shape the beam profile to a low variation top-hat profile. This DOE is used in series with a focusing lens to shape the beam to a desired diameter at the effective focal length (EFL). A beam profiler is used to measure and monitor the beam shape during each experiment.\\

The second section of the experiment is the launch package, which refers to the arrangement of the flyers and targets (Fig.~\ref{fig:1} F). Each launch package has lateral dimensions of 50 mm by 50 mm and contains an array of strategically spaced flyers: here, a 7 by 7 square array yielding 49 experiments per package. Multiple launch packages are fabricated in advance, and the spall experiments are performed systematically through the pre-fabricated launch packages to improve experimental throughput.  The design of the launch package is critical for consistently and reliably achieving planar impact and strong PDV return signals at high throughput rates.  It is a layered structure foil as shown in Fig.~\ref{fig:1} F and consists of a substrate, a flyer, a spacer, and a target. The glass substrates are 50 mm by 50 mm by 0.625 mm borosilicate glass from McMaster Carr, and the epoxy used for bonding is Henkel Loctite Ablestik 24. The flyers are 100 µm thick, 1.5mm diameter Aluminum from Alufoil, and the spacer is a 240 µm thick Kapton sheet with built-in double-sided silicone-based adhesive. 
The targets (samples) are prepared through double-sided polishing of larger area foils to a thickness of 200 µm +/- 10 µm, and then 3 mm disks are created using a TEM-punch \cite{mallick2021spall}. The flyer and target thickness are determined based on wave propagation analysis so that the spall plane occurs within the target. Their diameters are sufficiently large to avoid any 
unloading wave effects on the central area of interest during the time of interest. The flyers are pre-cut using a femtosecond laser to obtain a flyer with a sufficiently large diameter in order to prevent edge unloading from affecting the PDV signal and to guarantee impact planarity. The pulse laser is operated at $\sim$ 800 mJ and is focused to a spot size of 1.85 mm with a 250 mm EFL. This yields fluences of $\sim$ 30-32 J cm$^{-2}$ that drive impact velocities $\sim$ 550-600 ms$^{-1}$.\\

%The launch process proceeds through a sequence of ablation, acceleration, and finally impact. After reaching the desired velocity, the flyer impacts the target, propagating planar shockwaves throughout the flyer and target to generate the spall failure process. The particle velocity at the rear surface of the target is measured using PDV.\\

\begin{comment}

Our current understanding of the launch process is that it proceeds through an ablation, acceleration, and impact sequence.  The laser is focused on the glass substrate of the launch package and transmits through to ablate the epoxy layer. The expanding hot gas from the ablation process, confined by the glass and spacer, accelerates the flyer. After reaching a desired constant velocity, the flyer impacts the target, propagating planar shockwaves throughout the flyer and target to generate the spall failure process. The shockwaves are measured on the target's rear free-surface using PDV. Importantly, the flyers are precut using a femtosecond laser to obtain large area flyers and impact planarity.The pulse laser is operated at $\sim$ 800 mJ and is focused to a spot size of 1.85 mm with a 250 mm EFL. This yields fluences $\sim$ 30-32 Jcm$^{−2}$ that drive impact velocities $\sim$ 550-600 ms$^{−1}$.\\
\end{comment}

The diagnostics consist of a high-speed camera and a PDV system.  The high-speed camera (Shimadzu HPV-X) operates at 10 million frames per second. It provides a side profile view of the launch package for qualitative information regarding the flyer and impact planarity and a macroscopic assessment of the developed spall damage. The PDV system measures the normal particle velocity of the rear free surface of the target \citep{mallick2019laser, dolan2020extreme}. It is a heterodyne system that consists of two 1550 nm centered fiber lasers, a seed signal laser directed towards the target, and a 2.3 GHz upshifted reference signal for mixing. The seed laser is focused on the backside of the sample with a spot size of 80 $\mu m$; the material response is averaged over this area, so this length scale must be sufficiently large relative to the relevant material length scales. During the experiment, the reflected light (return signal) is imparted with a frequency shift based on the particle velocity. The return signal is mixed with the reference signal to obtain the beat frequency, which is measured and recorded using a 16 GHz LeCroy Oscilloscope. The strength of the return signal, and therefore the quality of the spall signal, is a strong function of the target’s surface roughness and orientation. We use a custom-built alignment apparatus that allows fine orientation control of the launch package for optimizing the return signal. Samples are double-sided polished to high reflectivity with diamond lapping paper to a 1-micron mirror finish to further maximize the return signal. \\

\begin{comment}

In this work, the glass substrates are 50 mm by 50 mm by 0.625 mm borosilicate glass from McMaster Carr. The epoxy is Henkel Loctite Ablestik 24.  The flyers are 100 $\mu m$ thick and 1.5mm diameter Aluminum from Alufoil. The spacer is a 240 $\mu m$ thick Kapton sheet  with built-in double-sided silicone-based adhesive. The targets are prepared through double-sided polishing of larger area foils to a thickness of ~200 $\mu m$ +/- 10 $\mu m$ and then 3mm disks are created using a TEM-punch \citep{mallick2021spall}.  The flyer and target thickness are determined based on wave propagation analysis so that the spall plane occurs within the target. Their diameters are sufficiently large to avoid any unloading wave effects on the central area of interest. The pulse laser is operated at $\sim$800mJ and is focused to a spot size of 1.85 mm with a 250 mm EFL.  This yields fluences $\sim$30-32 Jcm$^{-2}$ that drive impact velocities $\sim$550-600 ms$^{-1}$.\\

\end{comment}

\section{Results and Discussion}
\label{sec:results and discussion}

%%%%%%%%%%%%%SUB-SECTION%%%%%%%%%%%%%%%%%%%%%%%%%%%%%%%%%%%%%
\subsection{Initial Microstructure Characterization Results} 
\vspace{3mm}
\begin{figure*}[hbt!]
    \centering
    \includegraphics[width=0.85\linewidth]{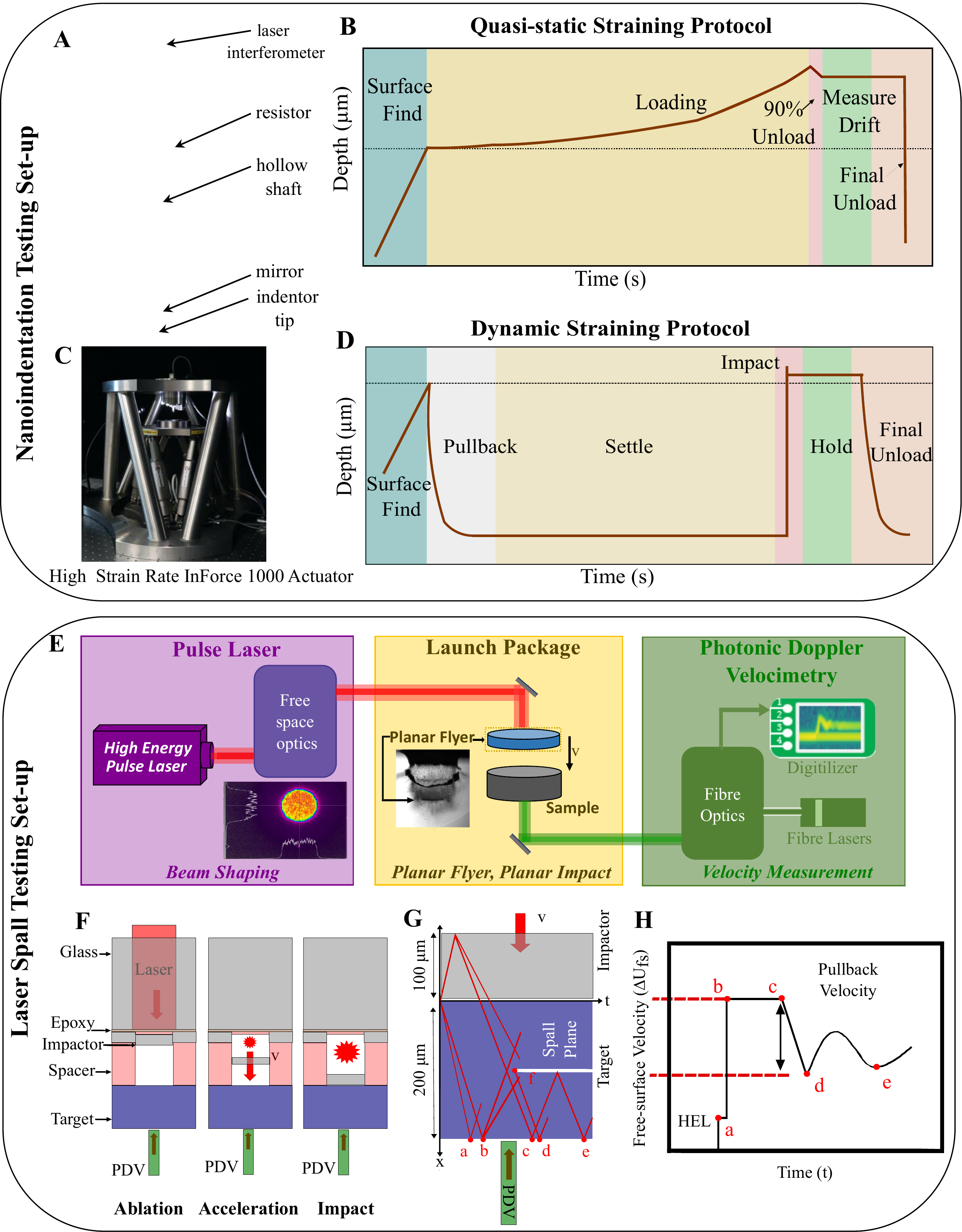}
    \caption{Schematic of nanoindentation setup (A and C) with typical depth vs. time plots for B) quasistatic and D) dynamic loading protocols. (E) Schematic of laser-driven micro-flyer shock setup with expected results collected to calculate spall strength. The overall layout of the testing system is broken into three major sections: the pulse laser, the launch package, and the photonic Doppler velocimetry system. (F) A side view schematic of the layered launch package depicting a typical single impactor-target configuration. This shows the three stages of the launch process: ablation of the epoxy, acceleration of the flyer, and finally, impact with the target. (G) A rotated view of a position vs. time Lagrangian diagram depicting the propagation of waves throughout the impactor and target that leads to the spall event, as well as the critical points measured via PDV. (H) An idealized velocity vs. time plot calculated from the PDV measurements on the target's free surface. The critical points correspond to the ones shown in (G).}
    \label{fig:1}
\end{figure*}

A Mg-5Zn (at\%) alloy, referred to as Z5 hereafter, was processed in two conditions: solutionized and peak-aged. EBSD of these two samples (Fig.~\ref{fig:2} A and C) showed the average grain size was around $\sim$ 205 $\mu$m for the Z5 solutionized (without precipitates) and $\sim$ 227 $\mu$m for the Z5 peak-aged sample (with precipitates). Furthermore, STEM micrographs (Fig.~\ref{fig:2} B and D) showed that Z5 solutionized was devoid of any precipitates within and along grain boundaries while the Z5 peak-aged sample had uniformly distributed precipitates from the aging treatment. The precipitate length is around $\sim$ 52 nm, and the areal density is $\sim$ 390 precipitates per $\mu m^2 $. Data from additional STEM TEM studies were used to verify the initial precipitate microstructures is shown in  \textcolor{blue}{\textit{SI Appendix}, FS.7 (a) to (b)}. 

%%%%%%%%%%%%%SUB-SECTION%%%%%%%%%%%%%%%%%%%%%%%%%%%%%%%%%%%%%
\subsection{Nanoindentation Results - Mechanical Property and Microscopy}
\vspace{3mm}
A custom nanoindentation protocol (Fig.~\ref{fig:1} A to D) was used to probe the mechanical properties at quasistatic and dynamic strain rate regimes. The indents across the strain rate regimes varied in size and are shown in Fig.~\ref{fig:3} A as a function of discrete strain rates. The nanoindentation hardness was measured between strain rates of ($10^{-2}$ to $10^{+4}$ s$^{-1}$) as shown in \textcolor{blue}{\textit{SI Appendix}, FS.3 (e) and (f)}. The hardness values were then converted to strength by a conversion factor of 1/3 \cite{bhattacharya1988finite, nix1998indentation}, and a graph of nanoindentation strength as a function of strain rate was plotted in (Fig.~\ref{fig:2} E and G). The depth vs. time and velocity vs. time profiles of nanoindentation experiments at both quasistatic and dynamic strain rate regimes are depicted in \textcolor{blue}{\textit{SI Appendix}, FS.1 (a) to (f)} and \textcolor{blue}{\textit{SI Appendix}, FS.2 (a) to (f)}, respectively. The corresponding strain rate and load profiles have also been indicated in these figure panels. Furthermore, we show the hardness vs. depth profiles for both Z5 solutionized and peak-aged at different strain rates in \textcolor{blue}{\textit{SI Appendix}, FS.3 (a) to (d)}. The average and standard deviation for hardness values were calculated between a depth of 2200 nm and 3000 nm. All data from these experiments are listed in \textcolor{blue}{\textit{SI Appendix}, Table S4-S6}.\\

The main takeaway from these results is that the Z5 peak-aged sample is stronger at the lower end of the strain rate regime (quasistatic) and the strength difference persists at the dynamic strain rate regime until $10^{+4}$/s.  The Z5 solutionized sample's average quasistatic strength is $\sim$ 254 MPa for strain rates ranging from $10^{-2}$ s$^{-1}$ to $10^{0}$ s$^{-1}$. The Z5 peak-aged sample's average quasistatic strength is  $\sim$ 301 MPa for the same range of strain rates. The dynamic strengths, though, increase with strain rate.  In the case of the Z5 solutionized sample, the average dynamic strength rises from $\sim$ 294 MPa at $10^{+1}$ s$^{-1}$ to $\sim$ 355 MPa at $10^{+4}$ s$^{-1}$. For the Z5 peak-aged sample, the average dynamic strength rises from $\sim$ 312 MPa at $10^{+1}$ s$^{-1}$ to $\sim$ 363 MPa at $10^{+4}$ s$^{-1}$.

%%%%%%%%%%%%%SUB-SECTION%%%%%%%%%%%%%%%%%%%%%%%%%%%%%%%%%%%%%

\subsection{Mechanisms Related to Quasistatic Nanoindentation}
\vspace{3mm}
At quasistatic and low values of strain rates ($10^{-2}$ s$^{-1}$ to $10^{0}$ s$^{-1}$), peak-aged Z5 is stronger than solutionized Z5. In the case of the peak-aged sample, finely distributed precipitates produced during the heat treatment process cover all the grains. When the indenter plunges into the peak-aged samples, the array of dislocations produced during deformation has to overcome these spatially distributed obstacles.   \\

The critical resolved shear stress (CRSS) for doing so can be predicted as $\Delta\tau_{Orowan}=G_mb/L$, where $G_m$ is the shear modulus of the matrix, $b$ is the Burgers vector, and $L$ is the spacing between precipitates \citep{Orowan1948symposium}. A further modification \cite{Hirsch1969} to capture the dependence of the geometric configuration of precipitates can be written as 

\begin{equation}
%\label{eq: 5}
    \Delta\tau _{Orowan} = \frac{G_m b}{2\pi (d_s - 2r_p)\sqrt{1-\nu}} \text{ln}\frac{2 r_p}{r_0},
    \label{OrowanEquation}
\end{equation}
Where $r_0$ is the core radius of the dislocation \citep{Kocks1975}, $r_p$ is the average radius of the precipitates on the slip plane,  $\nu$ is Poisson's ratio, $d_s$ is the spacing between the precipitates on the glide plane, and  $d_s=n_s^{-1/2}$ where $n_s$ denotes the number of precipitates per unit area. For the Mg system, one can calculate the effective planar, inter-precipitate spacing $\lambda _e=d_s-2r_p$ of different precipitate morphologies in the HCP system and estimate the change in $\tau _{Orowan}$ using
\cref{OrowanEquation}. Let $f$ denote the volume fraction of the precipitates, and assume that $r_0=b$. The Orowan CRSS from the $c$-axis precipitate rods present in the Mg-Zn alloy system is given by 

\begin{equation}
    \Delta\tau_{Orowan,c}=\frac{G_m b}{2\pi\left[\frac{0.953}{\sqrt{f}}-1\right]d_t \sqrt{1-\nu}}\text{ln}\frac{d_t}{b},
    \label{Eq:c-axisOrowan}
\end{equation}
where $d_t$ is the precipitate rod diameter \cite{prameela2022strengthening}. It follows from above that $\Delta\tau_{Orowan,c}$ depends strongly on $f$ and precipitate diameter ($d_t$), which in turn depend on the effective inter-particle spacing $\lambda_e$. From this equation, one can estimate the dependence of Orowon CRSS for any precipitate size and a given volume fraction \cite{prameela2022strengthening}. The Orowon increment in CRSS 
%yield stress 
from Eq. (\ref{Eq:c-axisOrowan}) for the peak aged Z5 alloy gives a value of $\Delta\tau_{Orowan,c}\sim 43$ MPa. \\ 

In the solutionized Z5, the alloying elements provide strengthening ($\Delta\tau_{ss}$) via distortion of the lattice, which may be estimated as 
\begin{equation}
    \Delta\tau_{ss}={B_r} c^{2/3} + {B_s} c^{2}\left(1-c\right)^2,
    \label{Eq:SS}
\end{equation}
where $c = 5\%$ denotes the nominal concentration of Zn, $B_r = 43.2$ MPa is the coefficient of random solid solution strengthening, and $B_s = 6$ GPa is the coefficient of strengthening associated with short-range order \cite{blake2008solid}. According to Eq. (\ref{Eq:SS}), $\Delta\tau_{ss}|_{c=5\%} \sim 19$ MPa, implying that the Zn alloying content is roughly twice as effective in precipitate form as compared to solution form. It is worth noting that peak-aged Z5 has $c \sim 1\%$ concentration of Zn that remains in solution, which results in a small solid solution strengthening of $\Delta\tau_{ss}|_{c=1\%} \sim 2$ MPa, according to Eq. (\ref{Eq:SS}). The difference in the the quasi-static yield strength $\Delta\sigma_Y$ can be estimated via the Taylor factor, i.e.\ $\Delta\sigma_Y = M_{P-A} \left(\Delta\tau_{Orowan,c} + \Delta\tau_{ss}|_{c=1\%}\right) - M_{S} \Delta\tau_{ss}|_{c=5\%}$, where $M_{S} \sim 4.5$ and $M_{P-A} \sim 3$ denote the Taylor factors for random solutionized Z5 and weakly textured peak-aged Z5, respectively. Substituting all the values, $\Delta\sigma_Y \sim 50$ MPa. This value is in good agreement with the experimentally measured difference in the average yield strength between the solutionized and peak-aged samples from nanoindentation experiments of $\sim$ 47 MPa and as shown in Fig.~\ref{fig:2} F.

%%%%%%%%%%%%%SUBSECTION%%%%%%%%%%%%%%%%%%%%%%%%%%%%%%%%%%%%%
\subsection{Mechanisms  Related to High Strain Rate Nanoindentation}
\vspace{3mm}

Zerilli and Armstrong~\cite{zerilli1996constitutive, zerilli1998dislocation} developed a constitutive model to characterize the strain-, strain rate- and temperature-dependent response of HCP metals, by combining the terms from their earlier BCC and FCC constitutive models~\cite{zerilli1987dislocation}. Specifically, Zerilli and Armstrong concluded that overcoming Peierls–Nabarro barriers, associated with dislocation motion, was the principal thermal activation mechanism for BCC metals, whereas dislocation interactions, and thus density, were the governing mechanism for FCC metals. Further, they considered HCP constitutive behavior to combine mechanisms of both BCC and FCC strain rate sensitivity. The current solutionized and peak-aged Z5 samples do, in fact, exhibit such a ``BCC response", that is, a rate-dependent initial yield strength commonly observed in BCC alloys. To this end, we employ the BCC term of the Zerilli-Armstrong \cite{zerilli1996constitutive} constitutive model for the yield strength, which is given by

\begin{equation}
    \sigma_Y = \sigma_G + \frac{k}{\sqrt{l}} + B \exp\left[-\beta_0 T + \beta_1 T \ln \frac{\dot\epsilon}{\dot\epsilon_0}\right],
    \label{ZAquation}
\end{equation}
where $\sigma_Y$ is the yield strength, $l$ is average grain diameter, $T$ is the absolute temperature, $\dot\epsilon$ is the strain rate, and $\dot\epsilon_0$ is the reference strain rate. The model parameters are $\sigma_G$, the quasi-static strength due to alloying content and pre-existing dislocations; $k$, the Hall–Petch slope; $B$, the strain-rate hardening modulus; $\beta_0$, the thermal-softening coefficient; and $\beta_1$, the rate-sensitivity parameter. \\

\begin{figure*}[hbt!]
    \centering
    \includegraphics[width=180mm]{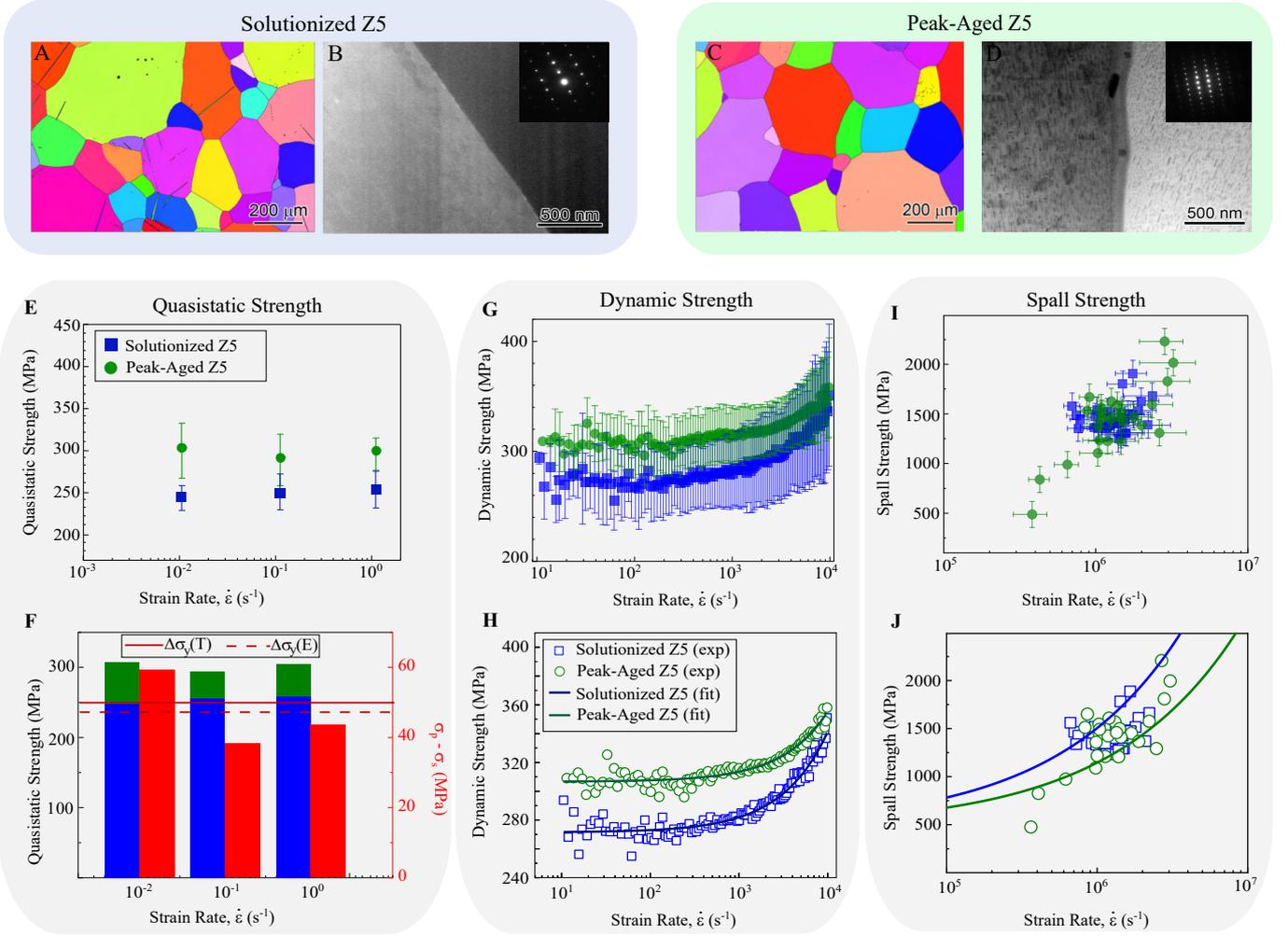}
    \caption{A) EBSD map of Z5 solutionized alloy at $150^{\circ}$C. B) STEM images of Z5 solutionized alloy at $150^{\circ}$C. C) EBSD map of Z5 peak-aged at $150^{\circ}$C. D) STEM images of  Z5 peak-aged at $150^{\circ}$C. E) Quasistatic strength measured via nanoindentation with error bars. F) Theoretical and experimental values of strength at quasistatic strain rates. G) Dynamic strength measured via nanoindentation. H) Theoretical dynamic strength predictions using the Zerilli-Armstrong model I) Spall strength measured via laser spall with uncertainty of spall strength parallel to the y-axis and uncertainty of strain rate parallel to the x-axis. J) Spall strength predictions using the Wilkerson-Ramesh model   }
    \label{fig:2}
\end{figure*}

The constitutive parameters required by the Zerilli–Armstrong model (Eq.~\ref{ZAquation}) were fit to the experimental results of dynamic strength using a nonlinear regression algorithm, and are reported in \textcolor{blue}{\textit{SI Appendix}, Table S1}. %With the choices $T=300$~K, $\beta_0=0$, and $k=526~\text{MPa}\cdot\mu\text{m}^2$~\cite{wang2012influence} for both Z5 samples, and $l=205~\mu\text{m}$ and $227~\mu\text{m}$ for solutionized and peak-aged samples, respectively, the predicted yield strengths as a function of strain rate are shown in Fig.~\ref{fig:2} H. 
The excellent agreement between the Zerilli-Armstrong model and the experimental data is shown in Fig.~\ref{fig:2} H, in support of the ``BCC response" assumption made in the constitutive analysis. \\
%Moreover, the constitutive parameters are summarized in \textcolor{blue}{\textit{SI Appendix}, Table S1}. 
%It might be seen that 

As expected, $\sigma_G$ of peak-aged Z5 is greater than solutionized Z5, because the Zn alloying content is more effective at strengthening in precipitate form as compared to solute form. Both peak-aged and solutionized Z5 are found to have the same $\beta_0$ and $\beta_1$. That said, the strain rate hardening modulus $B$ is found to be larger in solutionized Z5 as compared to peak-aged Z5. Given that $B$ is correlated with viscosity, this finding would seem to indicate that the viscosity of peak-aged Z5 is lower than the solutionized Z5. One possible explanation for this is that significant dislocation bowing around precipitates in peak-aged Z5 results in rapid dislocation multiplication at early deformation. The higher mobile dislocation density then results in a lower viscosity, \cite{nguyen2020physics}. Another contributing factor is that solute atoms often decrease the mobility of dislocations (and hence increase the overall viscosity), \cite{yi2016atomistic}. This argument would also support our finding that solutionized Z5 seems to be more rate-sensitive than peak-aged Z5.   %In addition, the peak-aged sample have a larger $\sigma_G$ than the solutionized one. This means that the precipitates govern initial strength, which is consistent with the results via quasistatic nanoindentation experiments. 
%However, regarding the strain rate hardening modulus $B$, the solutionized sample has a larger value, which indicates the solute left plays a more significant role on the hardening process. 
As a result of this higher rate sensitivity, the dynamic strength for these two alloys diminishes with increasing strain rate. Extrapolating our calibrated Zerilli-Armstrong model, we expect the solutionized Z5 to exhibit higher strength than peak-aged Z5 at sufficiently high strain rates, e.g. $\gtrsim 10^5$ s$^{-1}$. \\

In addition to the Zerilli-Armstrong model, we also employed three other commonly used constitutive models to describe the dynamic strengths of the solutionized and peak-aged Z5, i.e., the standard Johnson-Cook model~\cite{johnson1983constitutive}, the modified Johnson-Cook model with quadratic form proposed by Huh and Kang~\cite{huh2002crash}, and the Cowper-Symonds model~\cite{cowper1957strain} (see \textcolor{blue}{\textit{SI Appendix}, FS.4 (a) to (d)}). Comparing the coefficients of determination $R^2$ of these fitting results, we find that both the Zerilli–Armstrong and the Cowper-Symonds models provide the best prediction for the dynamic strengths depending upon a broad range of strain rates. A further check on these two models shows that they essentially have the same mathematical formulation, both expressing the yield strength as a power function of the strain rate. \\

Very few studies have reported high strain rate nanoindentation of Mg alloys, especially in the $10^{+3}$ s$^{-1}$ to $10^{+7}$ s$^{-1}$ regime. Researchers have tested pure Mg and dilute Mg alloys with nearly $ \sim$ 2-3 $\mu m$  grain sizes up to a $10^{+2}$ s$^{-1}$ strain rate \cite{somekawa2012high}. The small grain sizes ensured that the primary deformation mode was dislocation glide rather than twinning \cite{somekawa2012high}. The strength at high strain rates was influenced by the solute type, consistent with the theories of solid solution strengthening. Another study tested much coarser grained  Mg alloys with $ \sim$ 100 $\mu m$  grain sizes at strain rates up to $10^{+2}$ s$^{-1}$ \cite{somekawa2013nanoindentation}. In this case, twins were found to form during the early stages of nanoindentation, resulting in strain compatibility constraints leading to cross-slip promotion within the characteristic activation volume. Another study found that the yield point in nanoscale indents was often identified by pop-in events, which had a strong rate dependence and much lower activation volume \cite{somekawa2011effect}. In our study, given the large grain sizes, it is reasonable to assume that twinning is activated during the early stages of deformation indentation. Since the kinetics of twins and dislocations are closely related, using a pseudo-slip approach to model twins is reasonable. Therefore, the Zerilli-Armstrong model is appropriate for solutionized and peak-aged Z5 whose plasticity is mediated by either dislocations, twins, or a combination of both, although it was developed initially based on dislocation motion. Furthermore, the strain rate has a strong dependence on dislocation and twin mobility \cite{kannan2018mechanics}. Finely spaced precipitates are found to be effective obstacles to the motion of dislocations (affected by precipitate inter-spacing) and the motion of twins (affected by precipitate size) \cite{prameela2022strengthening}. Thus, these competing mechanisms play out in the case of the peak-aged Z5 sample, which has precipitates, as opposed to the solutionized Z5 sample with no precipitates. This further affirms the observation of the consistently higher dynamic strength of peak-aged Z5 over solutionized Z5 in the  $10^{+1}$ s$^{-1}$ to $10^{+4}$ s$^{-1}$ strain rate regime. 

\begin{figure*}[hbt!]
    \centering
    \includegraphics[width=0.9\linewidth]{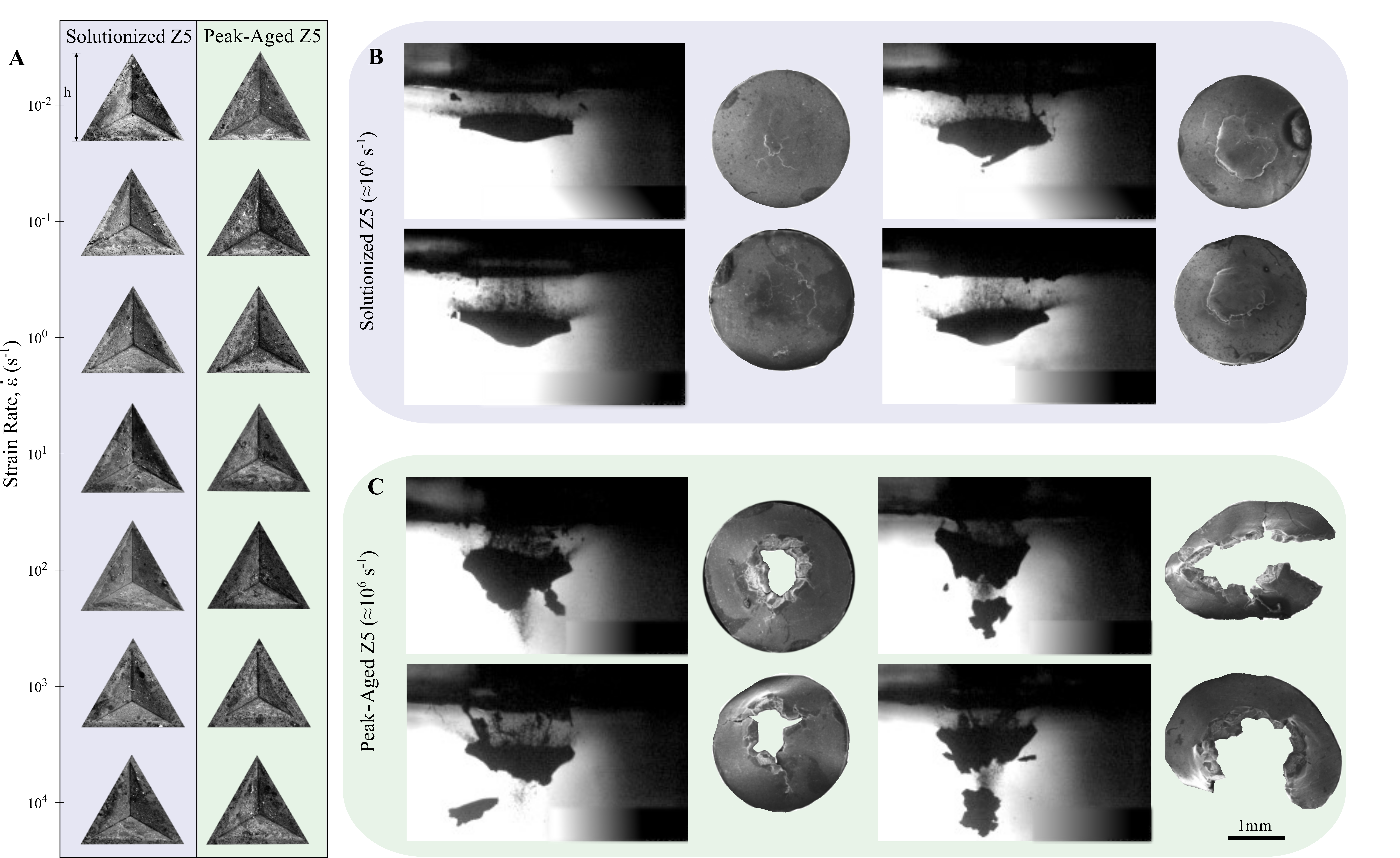}
    \caption{A) Nanoindentations of peak-aged and solutionized Z5 samples at quasistatic and dynamic strain rates. $h = 21 \;  \mu m$, $21 \; \mu m, 21 \; \mu m$, $59 \; \mu m$, $59 \; \mu m$, $29 \; \mu m$, $39 \; \mu m$ for each indentation at each strain rate, from low to high strain rates respectively. High-speed video shots and post-mortem SEM samples of B) Z5 solutionized, C) Z5 peak-aged after undergoing spall. The scale bar of 1 mm applies to all images shown in B and C.}
    \label{fig:3}
\end{figure*}

%%%%%%%%%%%%%SUB-SECTION%%%%%%%%%%%%%%%%%%%%%%%%%%%%%%%%%%%%%
\subsection{Spall Results - Mechanical Property and Microscopy} 
\vspace{3mm}
The fully spalled samples experience approximately planar separation within the material, parallel to the wavefront imposed by the flyer plate loading as a result of the dynamic tensile stresses in this plane. The initially compressive shock stress ($\Sigma_{S}$) generated in the target material from the loading is calculated from

\begin{equation}
    \Sigma_{S} = \dfrac{1}{2}\rho_0 U_{S} U_{B},
\label{equation:peak_shock}
\end{equation}

\noindent where $\rho_0$ is the reference density, $U_{S}$, is the shock speed estimated by assuming the linear equation of state with a parameter, $S_1$, of 1.21 from Marsh et al. \citep{marsh1980lasl} and $U_{B}$ is the maximum compression shock stress. The spall strengths of Z5 solutionized and Z5 peak-aged materials are obtained from the measured rear surface particle velocities using the following relationship: 

\begin{equation}
 \Sigma^* = \frac{1}{2}\rho_0 C_0 (\Delta U_{fs} + \delta),
 \end{equation} 

 where $\Sigma^*$ represents the spall strength, $\Delta U_{fs}$ is the velocity drop  seen in (Fig.~\ref{fig:1} H), $C_0$ is the bulk wave-speed  and $\delta$ elastic-plastic correction factor. The bulk wave speed was assumed to be 4540 m/s, the reference density as 1780 kg/m$^3$, and the elastic-plastic correction factor is zero  \cite{mallick2019laser}. The tensile strain rate, $\dot{\epsilon}$, is estimated via the velocity gradient and is given by

\begin{equation}
   \dot{\epsilon} = \dfrac{1}{2 C_{0}} \dfrac{\Delta U_{fs}}{|t_c - t_d|},
\end{equation}

\noindent where $t_c$ and $t_d$ are the times at points c and d in (Fig.~\ref{fig:1} H). Representative photon Doppler velocimetry spectrograms describing the time–frequency response of the spall signal are shown in Fig.~\ref{fig:1} G. More detailed spectrograms can be seen in \textcolor{blue}{\textit{SI Appendix}, FS.6 (a) and (b)} along with a compilation of the free-surface velocity traces for all spall experiments. The calculated spall strength and strain rate values are plotted in Fig.~\ref{fig:2} I. The tensile strain rates for both sample sets range from 10$^{+5}$ s$^{-1}$ to 10$^{+7}$ s$^{-1}$. The Z5 solutionized samples have an average spall strength of 1.44 $\pm$ 0.14 GPa, while the Z5 peak-aged samples have an average spall strength of 1.41 $\pm$ 0.35 GPa. Supplementary tables (\textcolor{blue}{\textit{SI Appendix}, TS.2 and TS.3}) provide a summary of sample thickness and spall results for the Z5 solutionized and Z5 peak-aged data sets, respectively. A t-test analysis suggests that there is no statistically significant difference in the spall strength between the two datasets, yielding a p-value of 0.77 (\textcolor{blue}{\textit{SI Appendix}, FS.5 (d)}). To limit the effect of strain rate dependency, which is amplified by the mismatched strain rate range between the data sets, we also perform the t-test on a subset of data, from 9.42x10$^{+5}$ s$^{-1}$ to 1.44x10$^{+6}$ s$^{-1}$, where both material preparations have a significant overlap in the volume of experiments. The same analyses show little change in mean and median in the subset of the experimental data. Representative images from high-speed photography for the spall experiments are shown in the left columns of Fig.~\ref{fig:3} B and C. The post-mortem SEM images of the 3 mm disks of both Z5 solutionized and Z5 peak-aged are shown in the right columns of Fig.~\ref{fig:3} B and C. For these spall experiments, the variation in $\Delta U_{fs}$ as a function of peak shock stress is shown in \textcolor{blue}{\textit{SI Appendix}, FS.5 (a)}.\\
 
 To see the trends in strength clearly, we also plot normalized effective stress (the ratio of shock stress to quasistatic yield strength) as a function of strain rate as shown in  \textcolor{blue}{\textit{SI Appendix}, FS.5 (b)} \cite{wu2003coupled}. The quasistatic yield strength used for this ratio was taken from the nanoindentation data at $10^{-2}$ s$^{-1}$. From analyzing the spall results, we see that the spall strengths are nominally the same between the solutionized and peak-aged samples, with the solutionized microstructure perhaps exhibiting a slightly higher spall strength. However, the fracture and subsequent failure of the samples from the spall are dramatically different. We see that the damage is much more significant in the case of the Z5 peak-aged samples, per SEM micrographs shown in Fig.~\ref{fig:3} B and C, likely from precipitate-mediated void nucleation and coalescence.  \\

The sharp contrast between the measured spall strength and the fracture morphology highlights the importance of not relying on the spall strength value alone to gain a complete understanding of the failure process of a material undergoing spall.  The large number of experiments in this work is necessary to confidently identify these trends that might otherwise be lost through traditional low-throughput methods. While the measured spall strengths are similar, the fracture surfaces are dramatically different. In high-strain rate applications, such as for protection materials, knowledge of both the developed stress state and the postmortem failure morphology, are crucial to understanding material failure.

%%%%%%%%%%%%%SUB-SECTION%%%%%%%%%%%%%%%%%%%%%%%%%%%%%%%%%%%%%

\subsection{Mechanisms Related to Spall} 
\vspace{3mm}
%\textcolor{red}{Writer: Debjoy}\\
%\textcolor{red}{Note from Suhas: I think we need to explain two things: a) basic spall stuff and b) why the peak-aged sample is failing more significantly than the solutionized sample. I have added a few equations below that may or may not be useful as you craft this section}

Our diagnostics provide both qualitative and quantitative understanding of the spall failure phenomena through imaging and in-situ velocimetry, respectively, yet the imaging techniques show the greatest difference in the failure process in our experiments. The high-speed video and postmortem microscopy, shown in Fig. \ref{fig:3} B and C, indicate that the spalled layer has completely fragmented away from the sample in the peak-aged Z5 case. In contrast, solutionized Z5 shows incomplete fragmentation in the spalled layer, more akin to a bulging separation. \\

%\citet{kanel1996spallation} present several contradictory spall experimental results where materials with the same spall strength show extremely varied modes of failure, from brittle single-crystal metals, to highly heterogenous ceramics, to polycrystalline metals. They emphasize that the spall strength measured from experiments like ours only defines conditions for damage nucleation and does not identify the mode of failure. In our experiments, the high-speed video and postmortem microscopy, shown in Fig. \ref{fig:3} B-C, indicate that the spalled layer has completely fragmenting away from the sample in the peak-aged Z5 case, In contrast, solutionized Z5 shows incomplete fragmentation in the spalled layer, more akin to a bulging separation. 

%Two treatments:

The observed dichotomy in the failure mechanism (Fig. \ref{fig:3}) is likely driven by defects or heterogeneities in the microstructure of the metallic alloy. Assuming that the damage during spall primarily nucleates (i) along grain boundaries or (ii) at precipitate-matrix interfaces, we can estimate differences in the density of sites for damage via void nucleation.  We assume that the critical pressure for void nucleation sites $N$ follows a bounded probability distribution function with a power-law exponent of $\beta$=3. Following the Wilkerson-Ramesh spall model \cite{wilkerson2016unraveling}, we assume that the density of potential nucleation sites along grain boundaries scales inversely with grain size $l$, and (following similar scaling arguments) that nucleation sites at second phase particles scale with the inverse cube of their mean spacing $d_s$, i.e. 

\begin{align}
    N(l,d_s)=N_1 \left(\frac{l_0}{l}\right)+N_2\left(\frac{d^0_s}{d_s}\right)^3,
\end{align}

\noindent where $N_1=1\:\mu\text{m}^{-3}$ and $N_2=10\:\mu\text{m}^{-3}$ are the densities of grain boundary and particle nucleation sites for a reference grain size of $l_0=1\:\mu\text{m}$, and reference mean particle spacing of $d^0_s=10$ nm, respectively. For peak-aged Z5, $l=227\:\mu\text{m}$ and the mean particle spacing is equal to the precipitate spacing, nominally $d_s=50$ nm. For solutionied Z5, $l=205\:\mu\text{m}$ and the mean particle spacing is taken to be a relatively large value governed by impurity content,  $d_s=1\:\mu\text{m}$. The lower bound of the probability distribution function for the critical nucleation pressure is assumed to be a third of the limit critical tensile pressure of an idealized elastic-perfectly plastic material containing an infinitely small pre-existing void, i.e.,

\begin{align}
    \mathcal{R}_y\equiv \frac{2}{3}\left(\sigma_G+\frac{k}{\sqrt{l}}\right)\left[1-\ln\frac{3}{2}\left(\frac{\sigma_G}{E}+\frac{k}{E\sqrt{l}}\right)\right],
\end{align}

\noindent with $E$= 47.4 GPa. The upper bound of the probability distribution function is taken as $\mathcal{R}_{eos}=$ 7 GPa, corresponding to the ideal spall strength of a perfect Mg crystal. The solid lines in Fig. \ref{fig:2} J are model predictions of spall strength according to the Wilkerson-Ramesh spall model \cite{wilkerson2016unraveling} invoking the aforementioned model parameters for solutionized and peak-aged Z5. Considering the experimental variability, the agreement between the model and experiments is remarkable.   \\

Supplementary figure (\textcolor{blue}{\textit{SI Appendix}, FS.5 (c)}) shows the theoretical predictions of the mean spacing between nucleated voids (dimples observed postmortem) on the spall surface of solutionized and peak-aged Z5 as a function of the experimentally measured spall strength. The dimples on the fracture surface of peak-aged Z5 are expected to be smaller by approximately a factor of 3 than for solutionized Z5. As such, the areal density of dimples on fracture surfaces is anticipated to be roughly a factor of 10 higher on peak-aged Z5 than solutionized Z5, so fractures linking the nucleated voids are expected to be significantly more prevalent in the peak-aged Z5. \\

While this model slightly over-predicts the difference in spall strength between the solutionized and peak-aged alloys, when compared to the averages from our laser-shock experiments, the model does offer a compelling explanation for the more brittle-like failure and complete separation of the spalled region, as observed in peak-aged Z5 (Fig. \ref{fig:3}). The model curves shown in (Fig. \ref{fig:2} J) employ a single set of nominal (average) values for precipitate spacing and grain boundary densities. As such, the model predictions are deterministic and make no attempt to capture the stochasticity observed in our dynamic strength or spall strength measurements (Fig. \ref{fig:2} H to J). The stochasticity is expected due to the fact that both nanoindentation and laser-driven micro-flyer plate experiments probe relatively local, relatively small volumes of material, which may not contain a statistically representative volume of microstructure. As such, the locally measured properties are themselves spatially dependent for any type of experiment that does not probe a statistically representative volume of material. Interestingly, peak-aged Z5 microstructure (Fig. \ref{fig:2} C and D) appears to exhibit significantly more spatial variability than the solutionized Z5 microstructure (Fig. \ref{fig:2} C and D), which would seem to suggest that it would exhibit greater spatial variability in properties. Indeed, our spall strength measurements (conducted at various locations in the target plate) show significantly higher (spatial) variability in the peak-aged Z5 as compared to the solutionized Z5 (\textcolor{blue}{\textit{SI Appendix}, FS.5 (d)}).

\section{Conclusions}
\label{sec:conclusion}

%%%%%%%%%%%%%SUB-SECTION%%%%%%%%%%%%%%%%%%%%%%%%%%%%%%%%%%%%%

In this work, we combined two powerful small-scale mechanical testing techniques, namely custom nanoindentation and laser-driven micro-flyer shock, to probe the mechanical properties of Mg alloys across a large strain rate regime ($10^{-2}$ s$^{-1}$ to $10^{+7}$ s$^{-1}$). We tested Z5 solutionized (no precipitates) and Z5 peak-aged (with precipitates) across quasistatic, dynamic, and spall regimes. We found that at low to medium strain rates (i.e., quasistatic and dynamic), the Z5 peak-aged sample had higher strength when compared to the Z5 solutionized sample. We measured Orowon yield stress increments at quasistatic strain rates and applied the Zerilli and Armstrong constitutive model to predict the dynamic strength values. At higher strain rates approaching $10^{+4}$ s$^{-1}$, the nano-indentation experiments and constitutive modeling showed converging strength for both peak aged and solutionized Z5. \\
%This was partly due to the precipitates inhibiting dislocation and twin motion and subsequent increment in the Orowon yield stress. However, the spall strength of these two samples was very similar at very high strain rates. 

We observed that spall strength, a value measured at ultra-high strain rates and captured in the initial moments of shock loading, was not influenced by the differences in void nucleation density introduced by precipitates in the peak-aged alloy. In contrast, our post-mortem observations of the spalled samples showed very different failure behavior between the solutionized and peak-aged samples. We adopted a void nucleation model to show the differences between solutionized and peak-aged Z5 samples. The Z5 peak-aged samples failed more significantly due to precipitate-mediated void nucleation and accelerated spall fracture. \\

This demonstrated the importance of not only relying on the measured physical quantity of spall strength, which is often the default in engineering design work at ultra-high strain rates, but also examining the damage morphology for a more complete understanding of material failure in extreme environments. Experiments on the same material systems through traditional low-throughput methods would not have readily shown these interesting trends. This study demonstrates the potential for using high-throughput techniques to quickly map the mechanical properties of various metallic alloys and aid in the Materials by Design paradigm. Finally, our study also highlights an important lesson in paying attention to how microstructures can fail differently despite having similar strength at very high strain rates. \\

%\begin{table}[h]

%\begin{tabular}{|c|c|c|c|c|c|}
%\begin{table}[h]
%\centering
%\caption{Precipitate area fraction of each precipitation mechanism at all hot compression temperatures, from TEM}
%\begin{tabular}{|c|c|c|c|c|c|}
%\hline
%\multirow{2}{*}{Temperature (°C)} & \multicolumn{5}{c|}{Total area fraction} \\ \cline{2-6} 
% & R1 & R2 & R3 & R4 & R5 \\ \hline
%25  & 0.0081 & 0.0127 & 0 & 0 & 0 \\ \hline
%100  & 0.1493 & 0 & 0.1549 & 0 & 0 \\ \hline
%150 & 0.0561 & 0.0509 & 0.0170 & 0.1068 & 0 \\ \hline
%200 & 0.1060 & 0.2631 & 2.5758 & 15.7982 & 1.0318 \\ \hline
%\end{tabular}
%\end{table}

\noindent \textbf{Data and code availability} The datasets generated and codes used in the current study are publicly available at \url{https://craedl.org/pubs?p=6352&t=3&c=187&s=hemi&d=https:\%2F\%2Ffs.craedl.org#publications}.

\section*{Acknowldegements}
The authors would like to gratefully acknowledge the financial and technical support from the Center for Materials under Extreme Dynamic Environment (CMEDE). The research was sponsored by the Army Research Laboratory and was accomplished under cooperative agreement numbers W911NF-12-2-0022 and W911NF-22-2-0014. The views and conclusions contained in this document are those of the authors and should not be interpreted as representing the official policies, either expressed or implied, of the Army Research Office or the U.S. Government. The U.S. Government is authorized to reproduce and distribute reprints for Government purposes, notwithstanding any copyright notation herein. The nanoindentation measurements were supported by the Department of Energy, National Nuclear Security Administration (NNSA), under award number DE-NA0003857.

\section{References}

\bibliography{mybibfile}

\onecolumn 

\section{Supplementary Information}

\subsection{Details regarding the data from custom Nanoindentation set-up}
\vspace{3mm}

\setcounter{figure}{0}
\renewcommand{\figurename}{FS}
\renewcommand{\thefigure}{\arabic{figure}}
\begin{figure}[H] 
    \centering
    \includegraphics[width=0.8\linewidth]{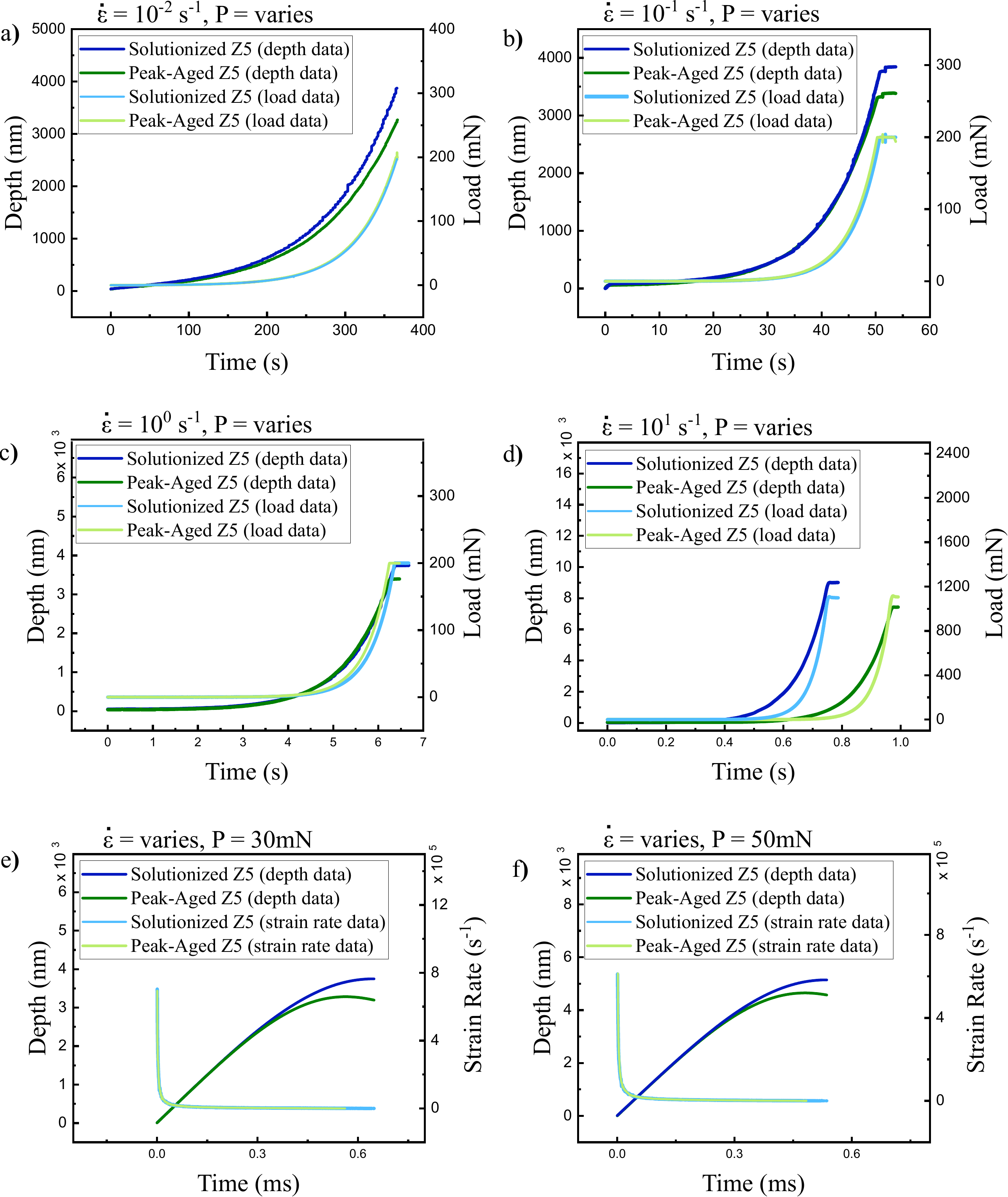}
    \caption{Depth vs. time for nanoindentation tests at different strain rates. For quasistatic tests (a-d), the load vs. time profiles are given to show the exponential loading used. For impact indentation tests (e-f), strain rate vs. time profiles are shown to indicate the non-constant strain rate during loading.}
    \label{fig:fs1}
   
\end{figure}

\begin{figure}[H] 
    \centering
    \includegraphics[width=0.8\linewidth]{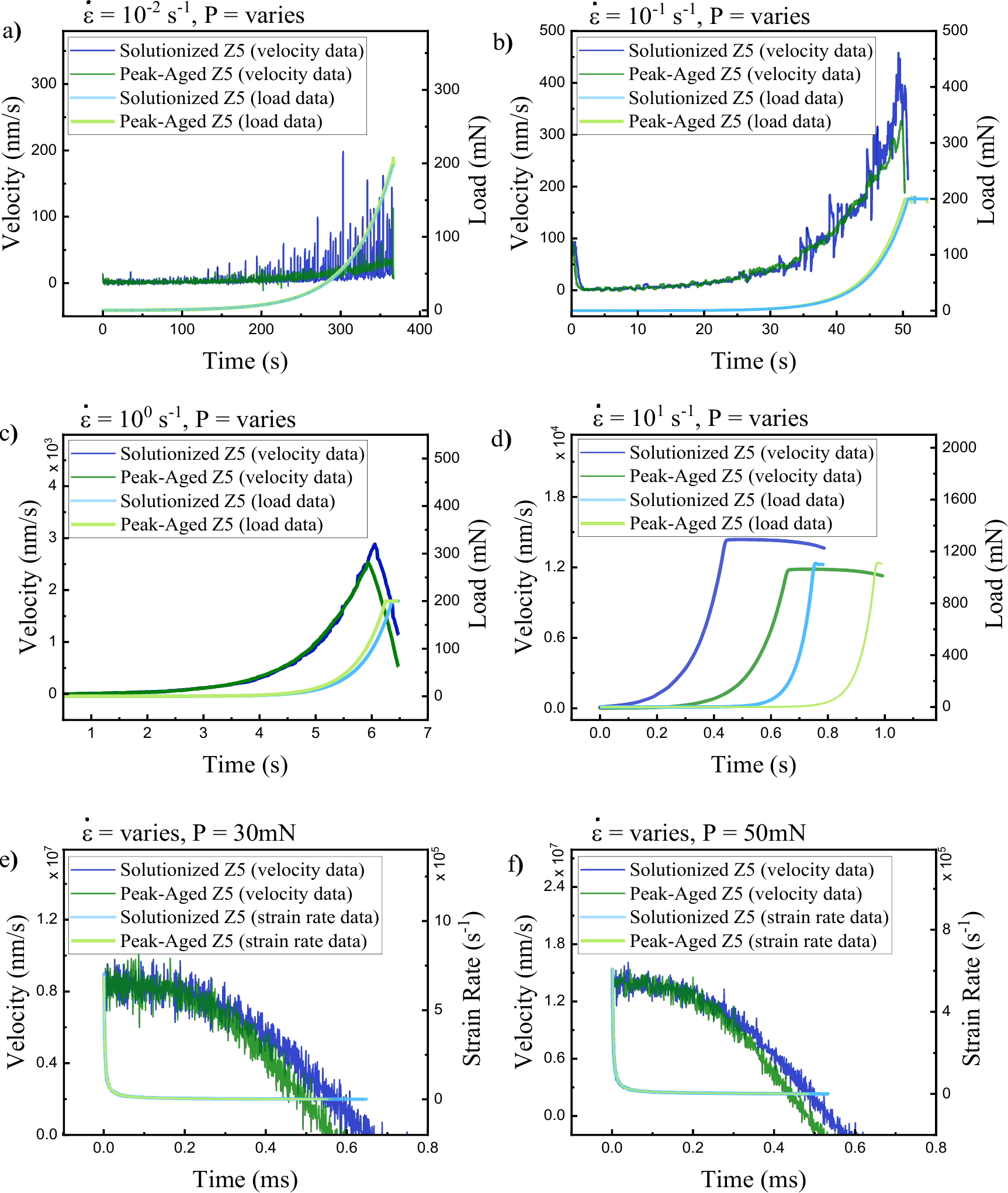}
    \caption{Velocity vs. time for nanoindentation tests at different strain rates. For quasistatic tests (a-d), the load vs. time profiles are given to show the exponential loading used. For impact indentation tests (e-f), strain rate vs. time profiles are shown to indicate the non-constant strain rate during loading.}
    \label{fig:fs2}
\end{figure}

FS \ref{fig:fs1} shows the exemplary depth-time and load-time data for the quasistatic nanoindentation testing. A \SI{1}{\kilo\hertz} feedback control loop was used to maintain a constant strain rate while loading to \SI{200}{\milli\newton}. The difference between the final depths of the Solutionized Z5 and Peak-Aged Z5 is a result of the small difference in hardness at each strain rate. The final load at $\dot\epsilon =$ \SI{E1}{\per\second} is much larger than the target \SI{200}{\milli\newton} due to dynamic effects brought on by the order of magnitude increase in velocity required to maintain the constant strain rate (seen in FS \ref{fig:fs2}).  Additionally, as the load is only updated in discrete steps every \SI{1}{\milli\second} by the control loop, the final load step is often larger than necessary to reach the target load in very fast tests.
FS \ref{fig:fs1} (e-f) shows depth-time and strain rate-time data for high strain rate impact nanoindentation tests. The strain rate is shown instead of load due to the strain rate not being constant throughout the experiment. Load is also not constant, but due to the nature of the impact test, a fixed impact force is given, and then most of the experiment is driven by dynamic forces from the rapid deceleration at the moment of impact.\\

FS \ref{fig:fs2} highlights velocity vs. time for each of the corresponding tests in FS \ref{fig:fs1}. The exponentially increasing velocity is necessary during quasistatic nanoindentation to maintain a constant strain rate as depth increases and, therefore contact area between the diamond tip and sample increases. The velocity at low strain rates (\SIrange{E-2}{E-1}{\per\second}) is particularly noisy due to the relatively small velocity of the test compared to the large $\Delta$velocity from the constant stiffness measurements (CSM) oscillation. At $\dot\epsilon =$ \SI{E0}{\per\second} (FS \ref{fig:fs2} (c)) the velocity of the test is much larger than the $\Delta$velocity created by CSM, making the noise negligible. At a strain rate of \SI{E1}{\per\second} (FS \ref{fig:fs2} (d)), CSM cannot be used since the high velocities result in a very short test and the \SI{1}{\kilo\hertz} control loop cannot produce the oscillatory frequency high enough to capture quality stiffness measurements while maintaining the increasing velocity. For \SIlist{E0; E1}{\per\second}, the decreasing and flat velocities (respectively) after reaching peak velocity are a result of the dynamic forces generated by the high velocities and accelerations and vary depending on the timing of the last force step and the amount of time it takes the feedback control loop to determine it is at or past the target load. For the impact tests in FS \ref{fig:fs2} (e-f), the test is designed so that contact with the sample surface is made at the instance of max velocity. No additional force is added to the system after this point to allow dynamic effects to drive the loading process. This results in a decreasing velocity throughout the entire experiment.\\

\begin{figure}[H] 
    \centering
    \includegraphics[width=0.5\linewidth]{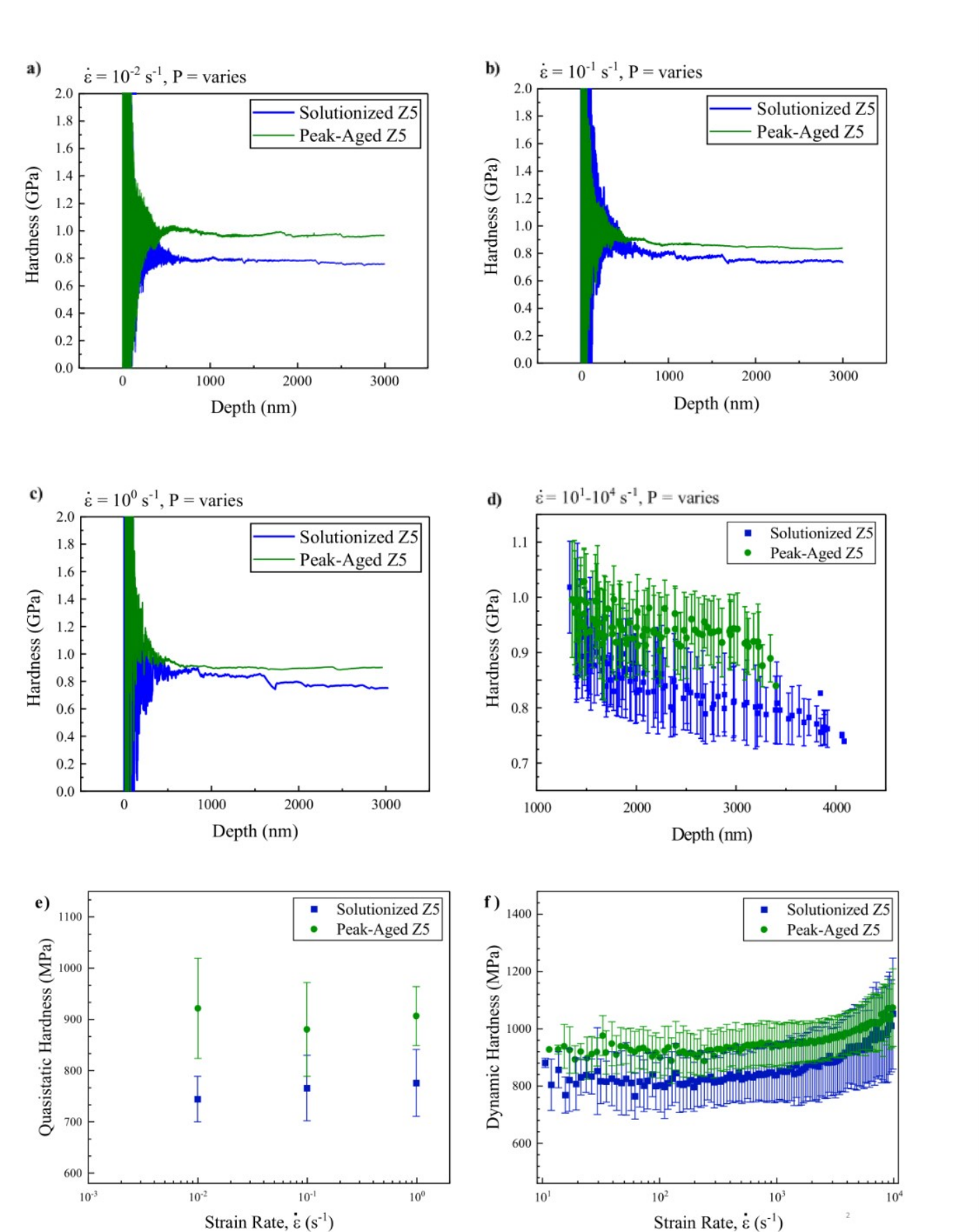}
    \caption{The average and standard deviation for hardness values were calculated between a depth of \SIlist{2200;3000}{\nano\meter}. A representative hardness vs. depth for nanoindentation tests for Z5 (wt\%) solutionized and peak-aged at strain rate a) $10^{-2} s^{-1} $, b) $10^{-1} s^{-1}$, c) $ 10^{0} s^{-1} $and d) $10^{1}$ to $10^{4} s^{-1}$, e) Quasistatic hardness measured via nanoindentation, f) Dynamic hardness measured via nanoindentation.}
    \label{fig:fs3}
\end{figure}

The measured hardness from both quasistatic and impact nanoindentation experiments are shown in FS \ref{fig:fs3}. FS \ref{fig:fs3} (a-c) shows that hardness vs. depth is fairly constant for both the solutionized Z5 and peak-aged Z5 samples after some initial indentation size effect (ISE). To mitigate the effects of ISE on the reported hardness, an average hardness between \SIlist{2200; 3000}{\nano\meter} is reported in FS \ref{fig:fs3} (e). Hardness vs. depth is also reported for the impact nanoindentation experiments in FS \ref{fig:fs3} (d), which shows a decreasing hardness as a function of depth even past a depth where hardness is flat in the quasistatic results. This decrease in hardness is explained by FS \ref{fig:fs3} (f), which shows hardness decreases as the strain rate decreases. During a nanoindentation impact experiment, the strain rate is not constant and decreases until the very end of the test where it falls to effectively \SI{0}{\per\second} at the peak load. Though strain rates as high as \SI{5E5}{\per\second} can be seen in FS \ref{fig:fs1} (e-f) and FS \ref{fig:fs2} (e-f),  hardness is only reported for $\dot\epsilon \le$ \SI{E4}{\per\second} to avoid edge effects created the strain rate being infinite at the moment of impact. \\

%%%%%%%%%%%%%%%%%%%%%%%%NEW-SECTION%%%%%%%%%%%%%%%%%%%%%%%%%%%%%%%%%%

\subsection{Details regarding the strength modeling on the Nanoindentation data}
\vspace{3mm}
\begin{figure}[H] 
    \centering
    \includegraphics[width=\linewidth]{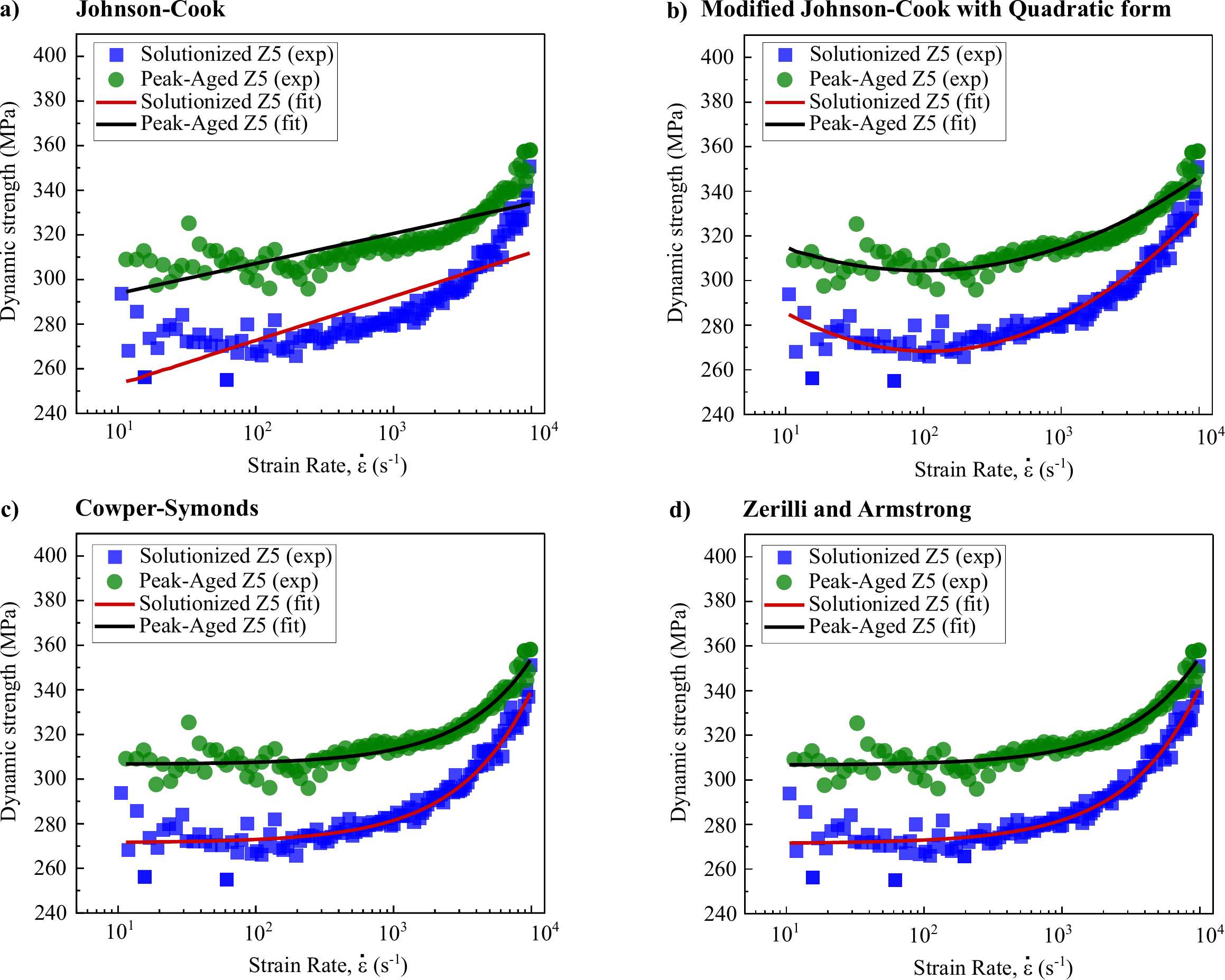}
    \caption{Comparisons of dynamic strengths between nanoindentation measurements and constitutive model fittings for solutionized and peak-aged Z5 samples: a) JC model with $R^2$ of $0.65$ for solutionized and $0.64$ for peak-aged, b) JC-quad model with $R^2$ of $0.92$ for solutionized and $0.88$ for peak-aged, c) CS model with $R^2$ of $0.94$ for solutionized and $0.91$ for peak-aged, and d) ZA model with $R^2$ of $0.94$ for solutionized and $0.91$ for peak-aged.}
    
    \label{fig:fs44}
\end{figure}

In addition to the Zerilli-Armstrong (ZA)  model, we also employ three other constitutive models to describe the dynamic strength of the solutionized and peak-aged Z5 depending upon strain rate, i.e., the standard Johnson-Cook (JC) model~\cite{johnson1983constitutive}, the modified JC model with quadratic form (JC-quad) proposed by Huh and Kang~\cite{huh2002crash}, and the Cowper-Symonds (CS) model~\cite{cowper1957strain}. Specifically, the JC form of the yield strength at room temperature is written as Eq~\cite{johnson1983constitutive}. In Eq \ref{Eq:JC}, $\epsilon_0$ is the reference strain rate. The fitting parameters are the quasi-static yield strength $\sigma_{Y0}$ and the strengthening coefficient of strain rate $C$. The JC-quad form of the yield strength is given by Eq \ref{Eq:JCquad}. In Eq \ref{Eq:JCquad}, $\sigma_{Y0}$, $C_1$ and $C_2$ are fitting parameters. Moreover, the yield strength with the CS form can be expressed by Eq \ref{Eq:CS}. In Eq \ref{Eq:CS} $\sigma_{Y0}$, $D$ and $P$ are fitting material constants. A comparison of the fitting results with experimental data is shown in FS.~\ref{fig:fs44}. The fitting parameters are tabulated in Table~\ref{tab:dynafit}.

%\begin{linenomath}
\begin{equation}
    \sigma_Y = \sigma_{Y0}\bigg(1+C\ln \Big(\frac{\dot\epsilon}{\dot\epsilon_0}\Big)\bigg),    
    \label{Eq:JC}
\end{equation}

\begin{equation}
    \sigma_Y = \sigma_{Y0}\bigg(1+C_1\ln \Big(\frac{\dot\epsilon}{\dot\epsilon_0}\Big)+C_2\ln \Big(\frac{\dot\epsilon}{\dot\epsilon_0}\Big)^2\bigg),
    \label{Eq:JCquad}
\end{equation}
\vspace*{-.1cm}
\begin{equation}
    \sigma_Y = \sigma_{Y0}\bigg(1+\Big(\frac{\dot\epsilon}{D}\Big)^{1/P}\bigg),
    \label{Eq:CS}
\end{equation}
\vspace*{-.1cm}
%\end{linenomath}

\begin{table}[H]
\centering
\caption{Constitutive model parameters for dynamic strengths of solutionized and peak-aged Z5.}
\begin{tabular}{|c|c|c|c|c|}
\hline
\multicolumn{2}{|c|}{Parameter} & Unit & Solutionized & Peak-aged \\
\hline
\multirow{2}{*}{JC model}
& $\sigma_{Y0}$ & MPa & $234.5$ & $280.4$ \\
& $C$ & - & $0.03606$ & $0.02084$  \\
\hline
\multirow{3}{*}{JC-quad model}
& $\sigma_{Y0}$ & MPa & $335$ & $348.5$ \\
& $C_1$ & - & $-0.08496$ & $-0.05405$  \\
& $C_2$ & - & $0.009067$ & $0.005781$  \\
\hline
\multirow{3}{*}{CS model}
& $\sigma_{Y0}$ & MPa & $271.5$ & $306.6 $ \\
& $D$ & - & $5.172\times10^4$ & $8.924\times10^4$  \\
& $P$ & - & $1.211$ & $1.172 $  \\
\hline
\multirow{3}{*}{ZA model}
& $\sigma_G$ & MPa & $234.74$ & $271.71$ \\
& $k$ & $\text{MPa} \cdot \mu \text{m}^{1/2}$ & $526$ & $526$ \\
& $B$ & MPa & $0.0347$ & $0.0182$  \\
& $\beta_0$ & $\text{K}^{-1}$ & $0$ & $0$  \\
& $\beta_1$ & $\text{K}^{-1}$ & $0.0028$ & $0.0028$  \\
\hline
\end{tabular}
\label{tab:dynafit}
\end{table}

%%%%%%%%%%%%%%%%%%%%%%%%NEW-SECTION%%%%%%%%%%%%%%%%%%%%%%%%%%%%%%%%%%
\subsection{Details regarding the data from custom laser spall set-up}
\vspace{3mm}

All spall data results, including raw and analyzed PDV results, high-speed camera footage, experimental parameter data sheets, and Python-based PDV code, are available via the link in the Data and Code Availability section of the main text. A reference file in the main directory summarizes the results and maps them with their relevant PDV and high-speed camera data files. A FileMaker relational database was created to track metadata associated with the experimental methods and techniques, and this data was exported into Excel sheets for reference.\\

Selected high-speed camera videos were chosen to clearly demonstrate the custom laser-driven micro-flyer experiment and the spall failure results of the two data sets. The videos were recorded from a side angle, showing the flyer accelerating from top to bottom in the frame. During impact experiments, the flyer is not visible to the PDV or camera. Therefore, the flyer's velocity and planarity are determined independently before impacting a sample. The first video demonstrates the launch of a single flyer in the absence of a sample to showcase the high degree of planarity achieved. The second and third videos show characteristic impact experiments of solutionized Z5 samples, while the fourth and fifth videos show characteristic impacts for peak-aged Z5 samples. The high planarity maintained during the impact event indicates the planarity of the event itself. Solutionized samples demonstrate higher damage resistance than peak-aged samples when impacted under the same conditions.\\

\vspace*{-5mm}

\begin{table}[H]
\centering
\begin{tabular}{|c|c|c|}
\hline
Video 1 & Flyer Launch & \href{https://www.dropbox.com/s/lwe828yffqp0ysf/Flyer-Camera_11_43_33-custom%201.mp4?dl=0}{\underline{Link 1}} \\
\hline
Video 2 & Solutionized Z5 Impact 1 & \href{https://www.dropbox.com/s/4oyi9atkqtq3kxf/Solutionized-Camera_14_46_16-custom%201.mp4?dl=0}{\underline{Link 2}} \\
\hline
Video 3 & Solutionized Z5 Impact 2 & \href{https://www.dropbox.com/s/r60674ad7m9x6i1/Solutionized-Camera_15_47_40-custom%201.mp4?dl=0}{\underline{Link 3}} \\
\hline
Video 4 & Peak-Aged Z5 Impact 1 & \href{https://www.dropbox.com/s/tiuha5u8nrhnni9/PeakAged-Camera_20_55_04-custom%201.mp4?dl=0}{\underline{Link 4}}  \\
\hline
Video 5 & Peak-Aged Z5 Impact 2 & \href{https://www.dropbox.com/s/512hv137cf3bf9h/PeakAged-Camera_19_54_34-custom%201.mp4?dl=0}{\underline{Link 5}} \\
\hline
\end{tabular}

\end{table}

\begin{figure}[H] 
    \centering
    \includegraphics[width=\linewidth]{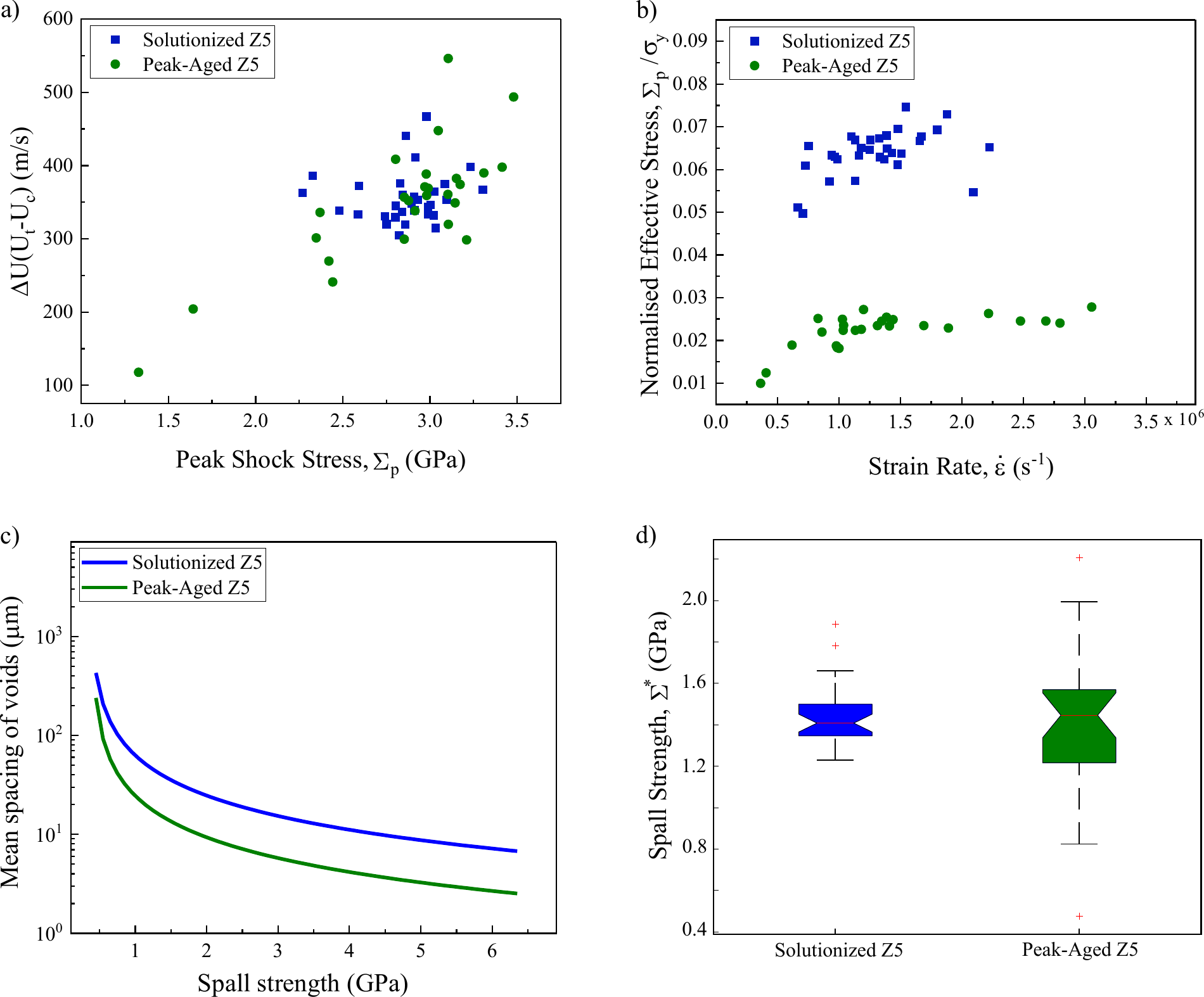}
    \caption{a)Delta U vs. Shock Stress, b)Normalized Effective Stress vs. Strain rate, c) Mean spacing of voids vs. experimentally measured spall strength, d) Spall strength of solutionized Z5 and peak-aged Z5 samples. }
    
    \label{fig:fs5}
\end{figure}

The spall results are presented in two plots in FS~\ref{fig:fs5}. FS~\ref{fig:fs5}a displays the pullback velocity against the peak shock stress for both sample sets. While spall results are typically plotted against strain rate, it is not directly controllable in laser-driven micro-flyer experiments. Instead, the impact velocity is controlled, which determines the peak shock stress in the material. As peak shock stress increases, metals experience strain-hardening from the initial shock compression wave before spall failure. The correlation is clearly observable with the peak-aged samples, while the solutionized sample set is too narrow to draw definitive conclusions. The box plot in FS~\ref{fig:fs5}c provides a more rigorous statistical analysis and a t-test. An ANOVA test indicates that there are no statistically significant differences in the spall strengths between the two datasets within a 95\% confidence interval, which is also evident from the comparison of the box-plot distributions. \\

%\textcolor{red}{*** PUT BOX PLOT HERE ***}

\begin{figure}[H] 
    \centering
    \includegraphics[width=0.7\linewidth]{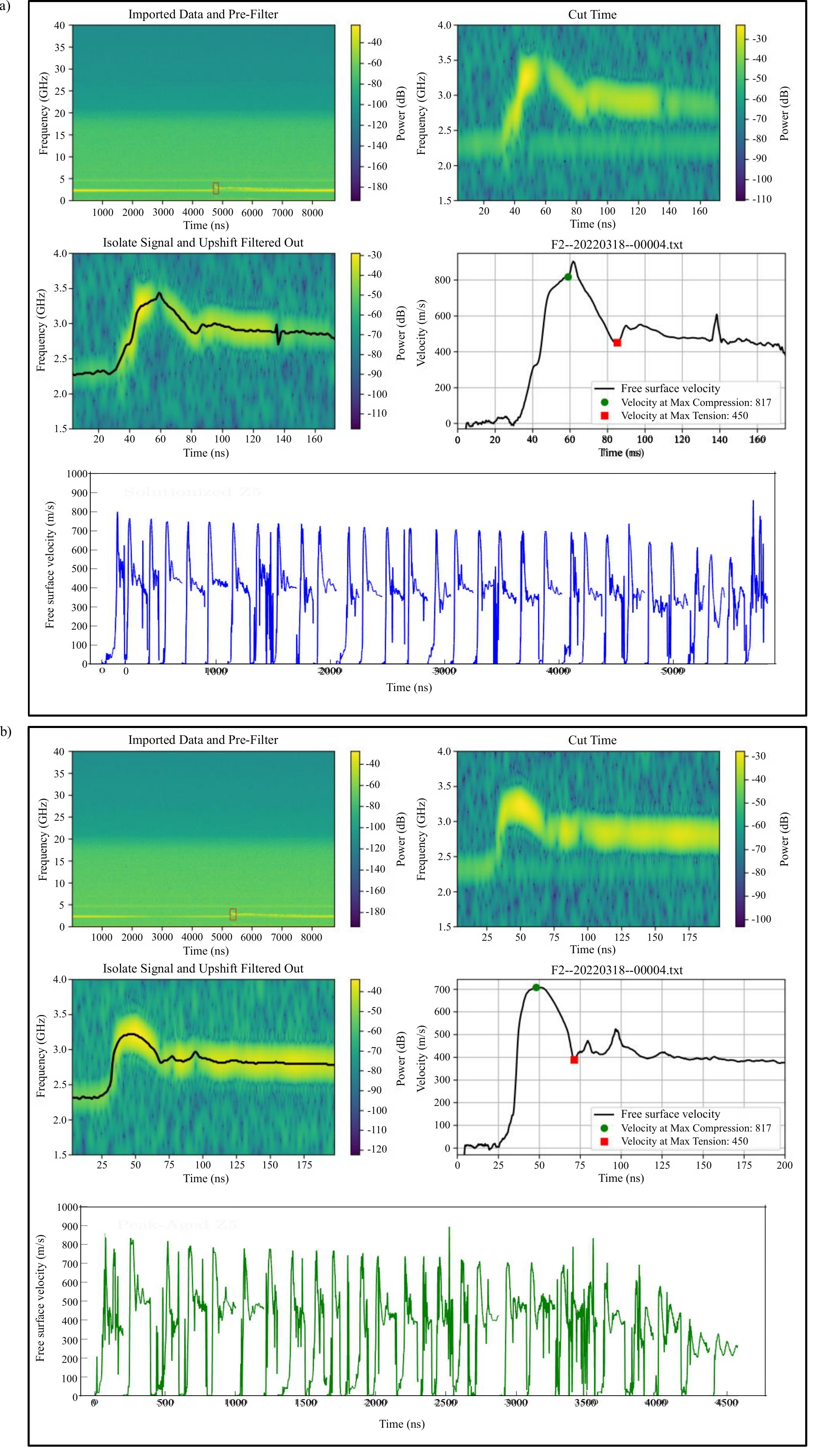}
    \caption{Representative photon Doppler velocimetry spectrograms describing the time-frequency response and free surface velocity of the spall signal from a) solutionized Z5 and b) peak-aged Z5}
    
    \label{fig:fs6}
\end{figure}

FS~\ref{fig:fs5}b plots the normalized effective stress against the strain rate, where the former is calculated as the peak shock stress divided by the yield strength. The yield strength of a material is a measure of its resistance to spall failure, with higher yield strength indicating higher resistance to void growth. In FS~\ref{fig:fs5}b, lower normalized effective stress corresponds to higher spall resistance. These results highlight the differences between the two datasets and are consistent with the observed variations in damage morphology. Specifically, the peak-aged samples exhibit a higher normalized effective stress and sustain more significant damage under the same impact conditions. FS~\ref{fig:fs6} shows a summary of the PDV results for both datasets, with essentially the same information but for each dataset displayed in the top and bottom frames. Within each frame, the top four figures represent how our PDV code extracts the PDV trace and relevant data points. While a short-time Fourier transform is used for viewing purposes, the PDV code employs direct phase differentiation for more accurate analysis of high-frequency waves. First, the spectrum is imported and coarsely pre-filtered to include only the expected data range. Second, the starting point of the signal is identified, and a finer time-based filter is applied based on the expected duration of the signal. Third, the spall signal is isolated by filtering out the frequency upshifted signal. Fourth, the signal is differentiated, the velocity trace is calculated, and the critical data points are automatically determined. The fourth frame shows the identification of the velocities at maximum compression and tension. The PDV code, processing input parameters, and a graphical summary of results are included in the shared data directory. Lastly, at the bottom of each frame, a compilation of all PDV traces is presented side-by-side for an easily viewed summary and direct comparison.

Tables~\ref{fig:tb5} and \ref{fig:tb6} provide a concise summary of the spall failure results for the solutionized and peak-aged samples, respectively. These tables include information such as the thickness of each individual sample, the peak shock stress, spall strength, and strain rate. The velocities at maximum compression and tension are also included in the shared data link. The thickness of each sample was measured using a micrometer prior to testing. The remaining quantities were determined based on the material density, the bulk wave speed, the equation of state, and the velocities at maximum compression and tension. These relationships are provided in the main text and were used to calculate the spall failure results for each sample.\\

FS \ref{fig:fs5} shows the theoretical predictions of the relationship between mean spacing of nucleated voids (dimples that might be observed postmortem) on the spall surface of the solutionized and peak-aged Z5 specimens as a function of the experimentally measured spall strength, per the analytic model by Wilkerson and Ramesh, 2016. In both the solutionized and peak-aged cases, the void spacing will decrease as the spall strength increases, but the increased density of failure nucleation sites in the peak-aged case will lead to smaller void spacing when compared to the solutionied case for a given spall strength, with the difference in spacing increasing as the spall strength increases.\\

\begin{comment}

\begin{figure}[H]
    \centering
    \includegraphics[width=0.5\linewidth]{2. Supplementary File/Figures/image.png}
    \caption{Theoretical predictions of the relationship between mean spacing of nucleated voids (dimples that might be observed postmortem) on the spall surface of the solutionized and peak-aged Z5 specimens as a function of the experimentally measured spall strength, per the analytic model by Wilkerson and Ramesh, 2016}
    
    \label{fig:fs55}
\end{figure} 

\begin{figure}[H]
    \centering
    \includegraphics[width=\linewidth]{2. Supplementary File/Figures/Figure-Mg-Zn-Processing-Effects-ANOVA.png}
    \caption{Caption}
    
    \label{fig:fs555}
\end{figure} 

\end{comment}

\begin{table}[H]
\centering
\caption{The results of spall experiments on solutionized Z5 alloy including thickness ($\mu$m), strain rate ($s^{-1}$), shock stress (GPa), spall strength (GPa) and pullback velocity (m/s).}
\begin{tabular}{|c|c|c|c|c|c|}
\hline
Shot No. & Thickness ($\mu$m) & Strain rate ($s^{-1}$) & Shock stress (GPa) & Spall strength (GPa) & Pullback (m/s) \\ \hline

1 & 214 & 1880410 & 3.58 & 1.61 & 398.47 \\
2 & 211 & 1658727 & 3.27 & 1.89 & 466.90 \\
3 & 207 & 2222355 & 3.20 & 1.66 & 411.15 \\
4 & 215 & 2093194 & 2.68 & 1.37 & 338.79 \\
5 & 213 & 1428659 & 3.13 & 1.78 & 440.73 \\
6 & 201 & 1367488 & 3.06 & 1.39 & 345.03 \\
7 & 269 & 1543896 & 3.66 & 1.48 & 366.94 \\
8 & 242 & 706241.6 & 2.44 & 1.47 & 362.80 \\
9 & 272 & 664606 & 2.51 & 1.56 & 386.22 \\
10 & 300 & 922959.1 & 2.81 & 1.35 & 333.23 \\
11 & 243 & 752198.5 & 3.21 & 1.43 & 352.94 \\
12 & 281 & 727316.8 & 2.99 & 1.34 & 330.86 \\
13 & 248 & 942287.7 & 3.11 & 1.46 & 360.31 \\
14 & 285 & 1163233 & 3.10 & 1.36 & 336.67 \\
15 & 244 & 1254023 & 3.28 & 1.39 & 344.02 \\
16 & 281 & 966358.6 & 3.09 & 1.52 & 375.55 \\
17 & 277 & 1180744 & 3.19 & 1.37 & 338.92 \\
18 & 259 & 1798959 & 3.40 & 1.52 & 375.08 \\
19 & 302 & 1506518 & 3.13 & 1.29 & 319.41 \\
20 & 248 & 1478650 & 3.41 & 1.43 & 352.94 \\
21 & 266 & 1332230 & 3.09 & 1.23 & 304.53 \\
22 & 283 & 985658.9 & 3.06 & 1.33 & 329.58 \\
23 & 270 & 1669877 & 3.33 & 1.47 & 364.66 \\
24 & 289 & 1132115 & 2.82 & 1.50 & 372.38 \\
25 & 232 & 1099777 & 3.32 & 1.34 & 332.03 \\
26 & 302 & 1250070 & 3.17 & 1.41 & 348.46 \\
27 & 278 & 1391273 & 3.19 & 1.44 & 357.35 \\
28 & 271 & 1476892 & 3.00 & 1.29 & 320.00 \\
29 & 242 & 1386178 & 3.33 & 1.27 & 314.82 \\
30 & 293 & 1328424 & 3.30 & 1.40 & 346.48 \\ 
31 & 242 & 1131126 & 3.28 & 1.35 & 333.15 \\ \hline
\label{fig:tb5}
\end{tabular}

\end{table}

\vspace{0.1cm}

\begin{table}[H]
\caption{The results of spall experiments on peak-aged Z5 alloy including thickness ($\mu$m), strain rate ($s^{-1}$), shock stress (GPa), spall strength (GPa) and pullback velocity (m/s).}

\begin{tabular}{|c|c|c|c|c|c|}

\hline
\centering

Shot No. & Thickness ($\mu$m) & Strain rate ($s^{-1}$) & Shock stress (GPa) & Spall strength (Gpa) & Pullback (m/s) \\ \hline

1 & 203 & 2217417 & 3.67 & 1.58 & 389.85 \\
2 & 203 & 1690616 & 3.27 & 1.45 & 359.02 \\
3 & 198 & 2798721 & 3.35 & 1.81 & 447.58 \\
4 & 207 & 2683489 & 3.42 & 2.21 & 546.11 \\
5 & 206 & 1890421 & 3.19 & 1.37 & 338.15 \\
6 & 206 & 3056042 & 3.88 & 1.99 & 493.58 \\
7 & 193 & 2476961 & 3.42 & 1.29 & 319.65 \\
8 & 213 & 830342.5 & 3.51 & 1.51 & 374.15 \\
9 & 185 & 1034565 & 3.12 & 1.44 & 356.61 \\
10 & 253 & 1182466 & 3.15 & 1.42 & 352.03 \\
11 & 249 & 1313375 & 3.27 & 1.57 & 388.32 \\
12 & 236 & 1028951 & 3.48 & 1.54 & 382.36 \\
13 & 214 & 1440590 & 3.47 & 1.41 & 348.92 \\
14 & 212 & 1133202 & 3.12 & 1.21 & 299.42 \\
15 & 200 & 986129.5 & 2.56 & 1.36 & 335.89 \\
16 & 203 & 979103.3 & 2.61 & 1.09 & 269.49 \\
17 & 236 & 407457.7 & 1.73 & 0.83 & 204.18 \\
18 & 197 & 618610.5 & 2.64 & 0.97 & 240.97 \\
19 & 203 & 363273.1 & 1.39 & 0.48 & 117.63 \\
20 & 199 & 1001282 & 2.53 & 1.22 & 301.16 \\
21 & 258 & 1038576 & 3.29 & 1.49 & 368.96 \\
22 & 207 & 862159.5 & 3.06 & 1.65 & 408.55 \\
23 & 216 & 1411300 & 3.26 & 1.50 & 370.82 \\
24 & 218 & 1347976 & 3.42 & 1.46 & 360.61 \\
25 & 236 & 1387231 & 3.55 & 1.21 & 298.53 \\
26 & 208 & 1200006 & 3.80 & 1.61 & 397.71 \\ \hline
 
\end{tabular}
\vspace{0.3cm}
\label{fig:tb6}
\end{table}

%%%%%%%%%%%%%%%%%%%%%%%%NEW-SECTION%%%%%%%%%%%%%%%%%%%%%%%%%%%%%%%%%%

\subsection{Additional Experimental Details}
\vspace{3mm}
\begin{comment}
   \textbf{XRD data for solutionized Z5 and peak aged Z5}
\begin{figure}[H]
    \centering
    \includegraphics[width=\linewidth]{2. Supplementary File/Figures/FS7 XRD ver 5.pdf}
    \caption{XRD of a) solutionized and b) peak-aged Z5 alloy.\cite{wang2012influence}}
    
    \label{fig:fs7}
\end{figure}  

\end{comment}

\begin{figure}[H] 
    \centering
    \includegraphics[width=\linewidth]{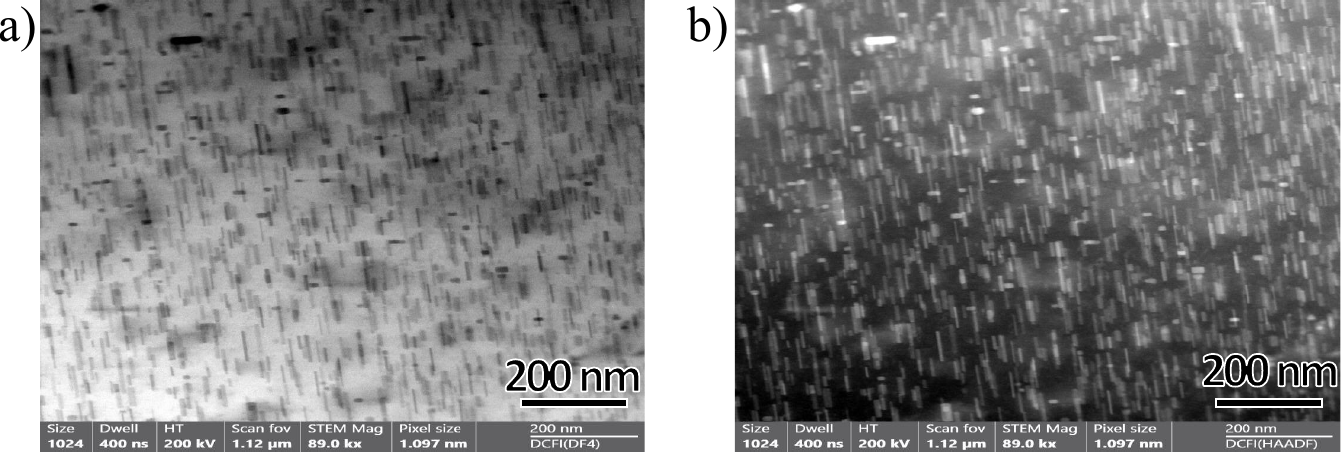}
    \caption{STEM images of a) Z5 peak aged observed under bright field microscope and b) Z5 peak aged observed under dark field microscope.}
    
    \label{fig:fs8}
\end{figure} 

\textbf{Standard deviation from the Nanoindentation experiments}

\begin{table}[H]
\centering
\caption{Standard deviation of Quasistatic hardness for solutionized Z5 and peak-aged Z5 samples.}
\begin{tabular}{|c|c|c|c|c|c|}
\hline
\multicolumn{3}{|c|}{Solutionized Z5} & \multicolumn{3}{|c|}{Peak-Aged Z5} \\
\hline
S.No & Strain Rate (1/s) & Quasistatic Hardness\_std & S.No & Strain Rate (1/s) & Quasistatic Hardness\_std \\
\hline
1 & 0.01 & 0.044385 & 1 & 0.01 & 0.097712 \\
2 & 0.1 & 0.063733 & 2 & 0.1 & 0.091532 \\
3 & 1 & 0.065156 & 3 & 1 & 0.057564\\
\hline
\end{tabular}
\end{table}

\begin{table}[H]
\centering
\footnotesize
\caption{Standard deviation of Dynamic hardness for Solutionized Z5 sample.}

\begin{tabular}{|c|c|c|c|c|c|}
\hline
S.No & Strain Rate (1/s) & Dynamic Hardness\_STD & S.No & Strain Rate (1/s) & Dynamic Hardness\_STD \\
\hline
1 & 10.5368 & 0.01615 & 61 & 1006.83789 & 0.09357 \\
2 & 11.87996 & 0.08946 & 62 & 1058.67786 & 0.10125 \\
3 & 13.71746 & 0.06307 & 63 & 1112.73409 & 0.10189 \\
4 & 15.63303 & 0.0629 & 64 & 1168.54589 & 0.10428 \\
5 & 16.9499 & 0.10939 & 65 & 1224.95179 & 0.10851 \\
6 & 19.38332 & 0.09094 & 66 & 1285.555 & 0.11345 \\
7 & 21.26067 & 0.06215 & 67 & 1345.54054 & 0.11029 \\
8 & 23.5918 & 0.06152 & 68 & 1409.77242 & 0.11027 \\
9 & 26.25462 & 0.06137 & 69 & 1479.21582 & 0.11207 \\
10 & 29.31402 & 0.15121 & 70 & 1550.52019 & 0.11519 \\
11 & 31.36326 & 0.07808 & 71 & 1621.00242 & 0.12046 \\
12 & 35.12035 & 0.07173 & 72 & 1692.96066 & 0.11825 \\
13 & 38.91717 & 0.09452 & 73 & 1769.79048 & 0.11893 \\
14 & 42.93214 & 0.10059 & 74 & 1850.61845 & 0.11277 \\
15 & 47.28611 & 0.11147 & 75 & 1935.71226 & 0.12544 \\
16 & 51.71428 & 0.12663 & 76 & 2020.78274 & 0.12857 \\
17 & 57.09975 & 0.09618 & 77 & 2109.09653 & 0.12327 \\
18 & 61.56953 & 0.07985 & 78 & 2201.42923 & 0.12732 \\
19 & 67.49483 & 0.08892 & 79 & 2298.89449 & 0.12484 \\
20 & 73.44082 & 0.09063 & 80 & 2394.89766 & 0.11944 \\
21 & 79.09578 & 0.09768 & 81 & 2495.77787 & 0.12956 \\
22 & 86.77363 & 0.07735 & 82 & 2605.74169 & 0.13207 \\
23 & 92.97928 & 0.08211 & 83 & 2711.33234 & 0.12492 \\
24 & 102.04513 & 0.09434 & 84 & 2824.85956 & 0.12661 \\
25 & 110.29707 & 0.11099 & 85 & 2941.44724 & 0.11752 \\
26 & 117.91848 & 0.09071 & 86 & 3061.39759 & 0.11652 \\
27 & 127.6346 & 0.09414 & 87 & 3182.59563 & 0.12828 \\
28 & 137.35956 & 0.10321 & 88 & 3315.34195 & 0.11445 \\
29 & 148.41189 & 0.09947 & 89 & 3443.04218 & 0.12567 \\
30 & 158.94798 & 0.09437 & 90 & 3582.2955 & 0.12406 \\
31 & 170.66804 & 0.08901 & 91 & 3718.89783 & 0.12712 \\
32 & 184.3826 & 0.10115 & 92 & 3869.18493 & 0.13725 \\
33 & 196.7869 & 0.08587 & 93 & 4017.25116 & 0.12485 \\
34 & 210.43677 & 0.08737 & 94 & 4168.54847 & 0.12696 \\
35 & 225.56907 & 0.09862 & 95 & 4331.43286 & 0.12761 \\
36 & 240.88302 & 0.09293 & 96 & 4493.13283 & 0.13195 \\
37 & 256.87634 & 0.07142 & 97 & 4660.26291 & 0.14771 \\
38 & 275.35135 & 0.0887 & 98 & 4837.76617 & 0.13474 \\
39 & 293.73678 & 0.08964 & 99 & 5017.62166 & 0.13879 \\
40 & 311.97289 & 0.08487 & 100 & 5189.481 & 0.13698 \\
41 & 331.23519 & 0.08369 & 101 & 5380.76496 & 0.14203 \\
42 & 352.6805 & 0.09294 & 102 & 5583.12942 & 0.13622 \\
43 & 376.23075 & 0.08993 & 103 & 5783.47385 & 0.14555 \\
44 & 398.6735 & 0.09228 & 104 & 5975.85655 & 0.14192 \\
45 & 423.94248 & 0.08565 & 105 & 6191.0202 & 0.13435 \\
46 & 449.21152 & 0.08802 & 106 & 6417.74146 & 0.14723 \\
47 & 475.82454 & 0.09796 & 107 & 6639.16311 & 0.16445 \\
48 & 504.91543 & 0.08404 & 108 & 6874.66909 & 0.15783 \\
49 & 533.0923 & 0.0842 & 109 & 7102.18887 & 0.16578 \\
50 & 564.68611 & 0.08471 & 110 & 7347.30777 & 0.15691 \\
51 & 596.11318 & 0.08891 & 111 & 7594.78445 & 0.13629 \\
52 & 629.66989 & 0.08598 & 112 & 7854.36082 & 0.1576 \\
53 & 665.20734 & 0.09349 & 113 & 8109.97729 & 0.16195 \\
54 & 702.49227 & 0.0911 & 114 & 8382.98127 & 0.16341 \\
55 & 740.72401 & 0.09057 & 115 & 8649.03082 & 0.16926 \\
56 & 780.88126 & 0.09311 & 116 & 8945.83198 & 0.18487 \\
57 & 822.83684 & 0.10031 & 117 & 9235.93133 & 0.17883 \\
58 & 865.98754 & 0.10064 & 118 & 9540.39072 & 0.16098 \\
59 & 911.66121 & 0.09852 & 119 & 9841.1845 & 0.19399 \\
60 & 959.20555 & 0.09646 &  &  & \\
\hline
\end{tabular}
\end{table}

\begin{table}[H]
\centering
\footnotesize
\caption{Standard deviation of Dynamic hardness for peak-aged Z5 sample.}
\begin{tabular}{|c|c|c|c|c|c|p{0.6\linewidth}|}
\hline
S.No & Strain Rate (1/s) & Dynamic Hardness\_STD & S.No & Strain Rate (1/s) & Dynamic Hardness\_STD \\
\hline
1 & 11.4317 & 0 & 60 & 9847.84811 & 0.13541 \\
2 & 13.6059 & 0.01506 & 61 & 959.09597 & 0.07525 \\
3 & 15.34068 & 0.07662 & 62 & 1006.28321 & 0.07614 \\
4 & 17.09732 & 0.08133 & 63 & 1057.7589 & 0.07772 \\
5 & 18.93417 & 0.04807 & 64 & 1111.99088 & 0.07028 \\
6 & 21.22634 & 0.02831 & 65 & 1169.34094 & 0.07438 \\
7 & 24.0556 & 0.07375 & 66 & 1224.38848 & 0.07151 \\
8 & 26.33735 & 0.0614 & 67 & 1284.33602 & 0.07956 \\
9 & 29.12948 & 0.04923 & 68 & 1346.84544 & 0.07384 \\
10 & 32.66838 & 0.06954 & 69 & 1413.14168 & 0.07135 \\
11 & 34.83625 & 0.07593 & 70 & 1480.21105 & 0.07279 \\
12 & 39.06283 & 0.07651 & 71 & 1547.48288 & 0.07005 \\
13 & 42.68435 & 0.06671 & 72 & 1619.76735 & 0.06744 \\
14 & 46.66146 & 0.06548 & 73 & 1697.59205 & 0.07034 \\
15 & 51.6477 & 0.061 & 74 & 1771.03775 & 0.07 \\
16 & 56.48525 & 0.07657 & 75 & 1850.10272 & 0.06852 \\
17 & 61.35494 & 0.07169 & 76 & 1936.11215 & 0.06958 \\
18 & 66.7823 & 0.06858 & 77 & 2021.17417 & 0.07083 \\
19 & 72.75592 & 0.07499 & 78 & 2108.78121 & 0.07107 \\
20 & 79.22108 & 0.07385 & 79 & 2199.38027 & 0.06533 \\
21 & 86.62227 & 0.07211 & 80 & 2296.34924 & 0.07429 \\
22 & 93.20207 & 0.06887 & 81 & 2398.49568 & 0.07144 \\
23 & 101.08401 & 0.07093 & 82 & 2494.59703 & 0.06693 \\
24 & 110.4806 & 0.08604 & 83 & 2603.17046 & 0.07155 \\
25 & 118.78563 & 0.07363 & 84 & 2711.5216 & 0.0709 \\
26 & 126.10944 & 0.08027 & 85 & 2822.9624 & 0.07366 \\
27 & 137.75045 & 0.08467 & 86 & 2941.70274 & 0.0729 \\
28 & 148.36915 & 0.06979 & 87 & 3058.78717 & 0.0659 \\
29 & 159.36976 & 0.08922 & 88 & 3184.64014 & 0.07176 \\
30 & 170.63119 & 0.08195 & 89 & 3312.5738 & 0.07171 \\
31 & 184.15915 & 0.06833 & 90 & 3443.66464 & 0.06909 \\
32 & 196.82152 & 0.06965 & 91 & 3578.73639 & 0.07519 \\
33 & 210.05017 & 0.0644 & 92 & 3722.1544 & 0.07387 \\
34 & 224.44571 & 0.07862 & 93 & 3864.67708 & 0.07607 \\
35 & 241.10177 & 0.05486 & 94 & 4013.35726 & 0.07265 \\
36 & 258.33593 & 0.07631 & 95 & 4174.19533 & 0.07723 \\
37 & 274.58269 & 0.06617 & 96 & 4323.27745 & 0.07867 \\
38 & 292.54554 & 0.07678 & 97 & 4490.96335 & 0.07297 \\
39 & 312.2643 & 0.06738 & 98 & 4661.79432 & 0.07392 \\
40 & 331.84418 & 0.0753 & 99 & 4826.88121 & 0.0823 \\
41 & 352.08983 & 0.06947 & 100 & 5002.43394 & 0.08408 \\
42 & 374.76696 & 0.07932 & 101 & 5190.53005 & 0.08859 \\
43 & 398.12235 & 0.06503 & 102 & 5389.91886 & 0.08471 \\
44 & 422.10922 & 0.06256 & 103 & 5577.82261 & 0.07853 \\
45 & 448.38299 & 0.08049 & 104 & 5782.94352 & 0.0954 \\
46 & 476.39404 & 0.07129 & 105 & 5984.84442 & 0.09328 \\
47 & 503.35696 & 0.07023 & 106 & 6191.55544 & 0.09271 \\
48 & 533.48216 & 0.07662 & 107 & 6410.68876 & 0.09653 \\
49 & 565.34942 & 0.07533 & 108 & 6645.10721 & 0.09031 \\
50 & 596.27527 & 0.07716 & 109 & 6868.25354 & 0.10244 \\
51 & 630.23991 & 0.07234 & 110 & 7099.88638 & 0.09946 \\
52 & 664.73547 & 0.07376 & 111 & 7352.04756 & 0.09845 \\
53 & 703.19504 & 0.07361 & 112 & 7596.21588 & 0.09558 \\
54 & 741.68297 & 0.07765 & 113 & 7850.56626 & 0.09473 \\
55 & 779.67713 & 0.07314 & 114 & 8104.09223 & 0.1263 \\
56 & 822.55709 & 0.08055 & 115 & 8385.46575 & 0.09898 \\
57 & 868.25771 & 0.07562 & 116 & 8665.08226 & 0.10278 \\
58 & 910.25145 & 0.06809 & 117 & 8942.38358 & 0.10007 \\
59 & 9530.94562 & 0.11103 & 118 & 9245.37519 & 0.10506\\
\hline
\end{tabular}
\end{table}

\section{Terminology Table}

\begin{table}[H] 
\centering
%\footnotesize
\begin{tabular}{|llll|} 
\hline
\multicolumn{4}{|l|}{\textbf{Terminology} } \\
&&&\\
$\theta$ & Angle &  $\tau_s$ & Resolved shear stress  \\ 
$A$ & Area & $G_m$ & Shear modulus\\
$C_0$ & Bulk wave-speed & $\gamma$ & Shear strain\\

$b$ & Burger’s vector & $\tau$ & Shear stress \\ 
$\dot{H}$ & Change in hardness over time & $\Sigma^*$ &  Spall strength \\
$\tau_r$ & Critical resolved shear stress &  $\epsilon$ & Strain  \\
$N_1$ & Density of grain boundary &  $n$ & Strain hardening exponent  \\
$N_2$ & Density of particle nucleation sites & $\dot{\epsilon}$ & Strain rate \\
$N$ & Density of Potential Nucleation sites & $B$ & Strain rate hardening modulus \\
$r_0$ & Dislocation core radius & $\tau_{ss}$ & Strengthening via distortion of lattice \\
$\rho$ & Dislocation density & $M_S$ & Taylor factors for random solutionized Z5\\
$\delta$ & Elastic-Plastic correction factor & $M_{P-A}$ & Taylor factors for weakly textured peak-aged Z5 \\
$G$ & Formation energy & $T$ & Temperature \\
$d_s$ & Glide plane spacing of precipitates & $\beta_0$ &  Thermal-softening coefficient \\
$l$ & Grain size & $U_{fs}$ & Velocity drop \\
$k$ &  Hall-petch slope & $V$ & Volume \\    
$H$ & Hardness & $\sigma_{Y}$ & Yield sterngth \\
$h$ & Indentation depth & $E$ & Young's modulus \\
$\dot{h}$ & Indentation depth over time &  \multicolumn{2}{l|}{\textbf{Alloy Designations (All at.\%)}} \\
$\dot\epsilon_i$ & Indentation strain rate & &  \\
$w$ & Interaction energy & Z$n$ & $n$\% Zn, rest Mg \\
$\lambda_e$ & Inter-particle spacing &  \multicolumn{2}{l|}{\textbf{Abbreviations}} \\
$P$ & Load  &  &\\
$\dot{P}$ & Loading rate  & ADF & Annular dark-field \\
$c$ & Nominal concentration & CSM & Continous Stiffness Measurement  \\
$I$ & Nucleation rate & CRSS & Critical Resolved Shear Stress \\
$r_p$ & Planar radius &  DOE & Diffractive Optical Element \\
$\nu$ & Poisson's ratio & EFL & Effective Focal Length \\
$d_t$ & Precipitate diameter ($=2r_p$) & EBSD & Electron Backscatter Diffraction  \\
$t_t$ & Precipitate thickness  & GP zone & Guinier–Preston zone \\

$f$ & Precipitate volume Fraction & HCP & Hexagonal close-packed \\
$\mathcal{R}_y$ & Probability distribution function & LDMF & Laser Driven Micro-Flyer \\
$\sigma_G $ & Quasi-static strength  &  LIPIT & Laser Induced Particle Impact Test \\
$r$ & Radial distance  & PDV & Photonic Doppler Velocimetry  \\
$R$ & Radius  & RMS & Root Mean Square  \\
$\beta_1$ & Rate-sensitivity parameter &  SEM & Scanning Electron Microscope\\
$\rho_0$ & Reference density & STEM & Scanning Transmission Electron   Microscopy\\
$l_0$ & Reference grain size & TEM & Transmission Electron Microscopy \\
$d^0_{s}$ & Reference mean particle spacing & XRD & X-ray Diffraction\\
$\dot{\epsilon_0}$ & Reference strain rate  &  &\\

\hline

\end{tabular}

\end{table}

\section{Supplementary References}

\begin{enumerate}

    \item  Gordon R Johnson. A constitutive model and data for materials subjected to large strains, high strain rates, and high temperatures. Proc. 7th Inf. Sympo. Ballistics, pages 541–547, 1983.
    
    \item  Hoon Huh and WJ Kang. Crash-worthiness assessment of thin-walled structures with the high-strength steel sheet. International Journal of Vehicle Design, 30(1-2):1–21, 2002.
    
    \item  G Rr Cowper and Paul Southworth Symonds. Strain-hardening and strain-rate effects in the impact loading of cantilever beams. Technical report, Brown Univ Providence Ri, 1957.
    
\end{enumerate}

\end{document}